\documentclass[prb,twocolumn,showpacs,superscriptaddress,reprint,aps]{revtex4-1}

\usepackage{amsmath, amsfonts, amssymb}
\usepackage{enumerate}
\usepackage{graphicx}
\usepackage{color}
\usepackage{paralist}

\newcommand{\eref}[1]{Eq.~(\ref{#1})}
\newcommand{\fref}[1]{Fig.~\ref{#1}}
\newcommand{\sref}[1]{Sec.~\ref{#1}}

\newcommand{\up}{\uparrow}
\newcommand{\dw}{\downarrow}
\newcommand{\e}{\mathrm{e}}

\newcommand{\si}{\hat{\sigma}_{0}}
\newcommand{\sx}{\hat{\sigma}_{1}}
\newcommand{\sy}{\hat{\sigma}_{2}}
\newcommand{\sz}{\hat{\sigma}_{3}}
\newcommand{\ti}{\hat{\tau}_{0}}

\newcommand{\tz}{\hat{\tau}_{3}}

\newcommand{\mbf}[1]{\mathbf{ #1 }}

\DeclareMathOperator{\real}{Re}
\DeclareMathOperator{\imag}{Im}
\DeclareMathOperator{\sgn}{sgn}

\begin{document}


\title{Superconducting proximity effect in three-dimensional topological insulators in the presence of a magnetic field}

\newcommand{\wurzburg}{Institute for Theoretical Physics and Astrophysics,
	University of W\"{u}rzburg, D-97074 W\"{u}rzburg, Germany}
\newcommand{\nagoya}{Department of Applied Physics, Nagoya University, Nagoya, 464-8603, Japan}
\newcommand{\tokyo}{Department of Applied Physics, University of Tokyo, Tokyo 113-8656, Japan}
\newcommand{\riken}{RIKEN Center for Emergent Matter Science (CEMS), Wako 351-0198, Japan}

\author{Pablo Burset}
\affiliation{\wurzburg}
\author{Bo Lu}
\affiliation{\nagoya}
\author{Grigory Tkachov}
\affiliation{\wurzburg}
\author{Yukio Tanaka}
\affiliation{\nagoya}
\author{Ewelina M. Hankiewicz}
\affiliation{\wurzburg}
\author{Bj\"orn Trauzettel}
\affiliation{\wurzburg}

\date{\today}



\begin{abstract}
The proximity induced pair potential in a topological insulator-superconductor hybrid features an interesting superposition of a conventional spin-singlet component from the superconductor and a spin-triplet one induced by the surface state of the topological insulator. 
This singlet-triplet superposition can be altered by the presence of a magnetic field. 
We study the interplay between topological order and superconducting correlations performing a symmetry analysis of the induced pair potential, using Green functions techniques to theoretically describe ballistic junctions between superconductors and topological insulators under magnetic fields. 
We relate a change in the conductance from a gapped profile into one with a zero-energy peak with the transition into a topologically nontrivial regime where the odd-frequency triplet pairing becomes the dominant component in the pair potential. 
The nontrivial regime, which provides a signature of odd-frequency triplet superconductivity, is reached for an out-of-plane effective magnetization with strength comparable to the chemical potential of the superconductor or for an in-plane one, parallel to the normal-superconductor interface, with strength of the order of the superconducting gap. 
Strikingly, in the latter case, a misalignment with the interface yields an asymmetry with the energy in the conductance unless the total contribution of the topological surface state is considered. 
\end{abstract}

\maketitle

\section{Introduction \label{sec:intro}}
A three-dimensional topological insulator (3DTI) is an extraordinary material with an insulating band gap but topologically protected gapless surface states. These states are possible due to spin-orbit interaction and follow a spin-polarized Dirac-type spectrum \cite{Qi_RMP}. 
Among many unique properties of these materials, one has triggered an intense interest in the condensed matter community: the emergence of Majorana fermions when TIs are in electrical contact with a superconductor \cite{Alicea_Majorana,*Beenakker_Majorana}. 
Indeed, the topological insulator-superconductor system can be used to engineer an effective \textit{spinless} triplet superconductor which is expected to host zero-energy topologically protected Majorana bound states \cite{Fu_2008,*Sau_2010,*Lutchyn_2010,*Alicea_2010,*Oreg_2010,*Potter_2010,*Potter_2011,*Prada_2012,*Houzet_2013,*Tiwari_2013,*Tiwari_2014b,*Rex_2014}. 


\begin{figure}[ht!]
	\includegraphics[width=\columnwidth]{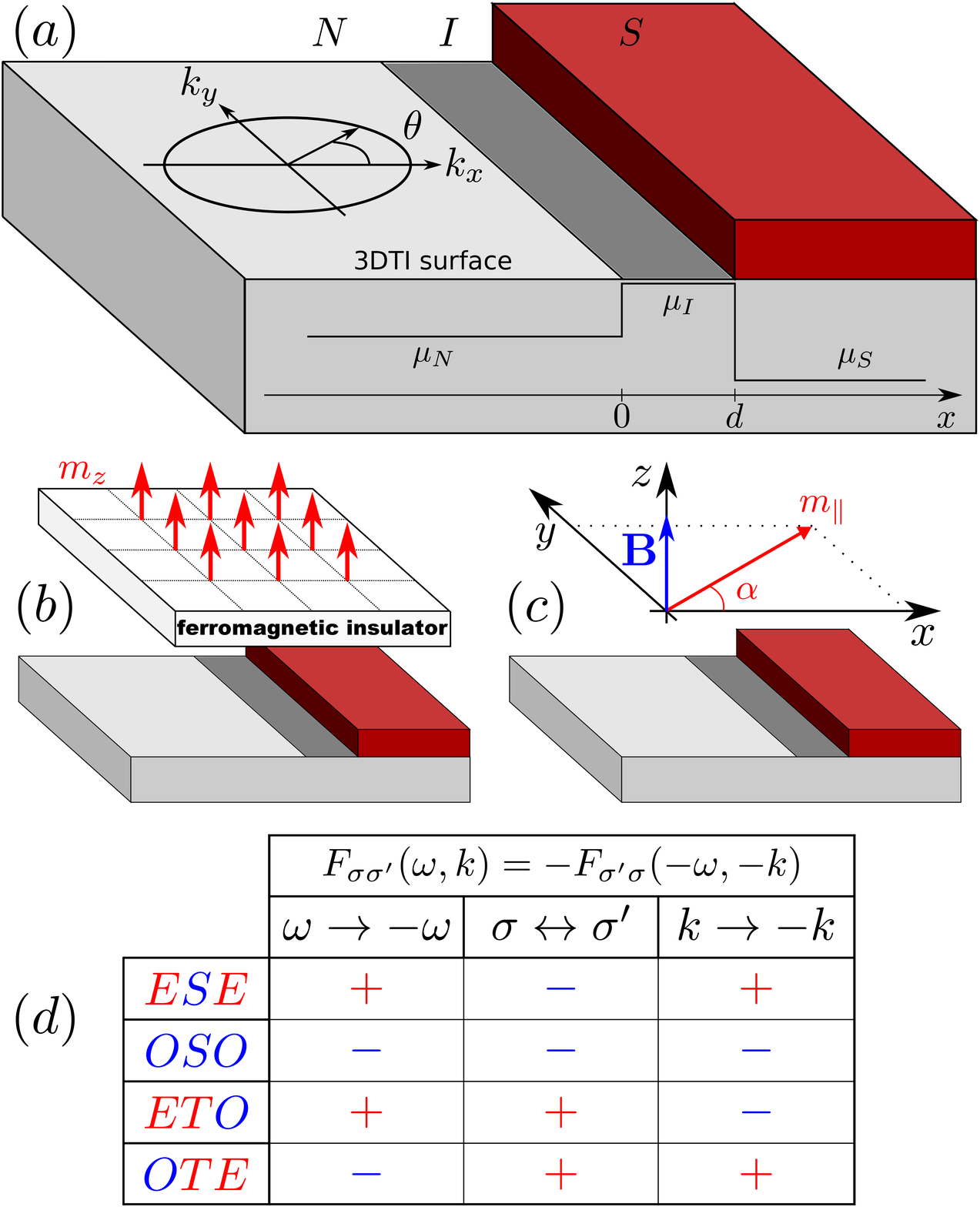}
	\caption{\label{fig:sketch}
		(a) Sketch of a normal-superconductor junction on the surface state of a topological insulator (regions $N$ and $S$, respectively) with an intermediate region $I$ modeled as a square potential barrier (see potential profile). The angle of incidence $\theta$ is measured with respect to the $x$-direction. 
		(b) An out-of-plane magnetization $m_z$ is induced in the system by proximity from a ferromagnetic insulator. 
		(c) An external magnetic field, stemming from the vector potential $\mbf{A}$ (blue arrow) with arbitrary orientation, also generates an effective magnetization (red arrows). (d) Symmetry classification of the anomalous Green function $F_{\sigma\sigma'}(\omega,\mbf{k})$. With respect to the exchange of the time (frequency), spin, and spatial (momentum) coordinates, the Green function can be even (E) or odd (O). The spin part can be divided into a singlet (S) or three triplet (T) components. The Pauli principle allows for four different combinations; using a ``frequency/spin/momentum'' notation: ESE, OSO, ETO, and OTE. }
\end{figure}

Remarkably, the interplay between the spin-momentum locking at the surface of the 3DTI, the conventional pair potential from a superconductor, and magnetic order from either a ferromagnet or a magnetic field creates a very exotic induced pair potential in the 3DTI. 
First, the proximity induced pair potential from a conventional singlet $s$-wave superconductor on the surface of a 3DTI features a superposition of singlet and triplet states \cite{Read_2000,Gorkov_2001,Schnyder_2008,Yokoyama_2012,Tkachov_2012,Tkachov_2013,Tkachov_2013b,Burset_2014}. Owing to the symmetry of Cooper pairs, the singlet (triplet) components are even (odd) under the exchange of the spatial coordinates of the two electrons (even/odd in momentum). At the same time, they are all even under the exchange of time coordinates (even in frequency). 
Second, at the interface between normal and superconducting regions inversion symmetry is broken. Such inhomogeneity generates new pairing components that are odd (even) in momentum if they are a singlet (triplet) state of spin \cite{Tanaka_2007,*Eschrig_2007,*Tanaka_2007b,*Tanaka_2007c,*Tanaka_JPSJ,Yokoyama_2012,Black-Schaffer_2012,*Black-Schaffer_2013,*Black-Schaffer_2013b}. Therefore, the Pauli principle imposes that these new terms must be odd in frequency. 
Finally, as we demonstrate in this paper, magnetic order directly affects all these components and opens the possibility for the exotic odd-frequency terms to become dominant and produce strong changes in transport observables of these systems. 

Among the unconventional components of the pair potential, a new, very rare type of superconducting condensate can be induced in this system: one with a spin-triplet component that is odd in frequency and even in momentum \cite{Berezinskii_1974,*Berezinskii_1974R,*Balatsky_1992}. 
Such exotic \textit{odd-frequency triplet pairing} has been proposed to explain the long-range proximity effect observed in superconductor-ferromagnet hybrids \cite{Bergeret_2001,*Birge_2010}. On such systems, the odd-frequency triplet component can be filtered from the rest of induced components of the superconducting condensate since it is the only one resistant to disorder \cite{[{For more information about odd-frequency triplet pairing on ferromagnet-superconductor hybrids see }][{}] Bergeret_RMP,*Eschrig_2010}. 
In topological systems, spin-polarization can be used to detect this unconventional pairing. Recently, Cr\'{e}pin \textit{et al.} have proposed a detection scheme based on nonlocal conductance measurements on the one-dimensional edge of a two-dimensional TI \cite{Crepin_2015}. 
Moreover, odd-frequency triplet pairing and the emergence of Majorana bound states are related: any system that supports Majorana zero-energy bound states also develops an odd-frequency spin-triplet component of the pair potential \cite{Asano_2013,Galitski_2014,DasSarma_2015,Ebisu_2015,Snelder_2015}. 

In this paper, we consider hybrid systems between 3DTI and superconductors like the one sketched in \fref{fig:sketch}(a). 
We study the proximity effect when such systems are exposed to a magnetization extended over the whole junction. The magnetization can be induced by proximity to a ferromagnetic material (b) or by an external magnetic field (c). 
Previous works on 3DTI-based normal-superconductor (NS) and Josephson (SNS) junctions focused only on transport signatures of such hybrids and the conditions for the formation of Majorana bound states \cite{Tanaka_2009,*Linder_2010b,*Sengupta_2013,Linder_2010,Olund_2012,Snelder_2013,Tkachov_2013,Tkachov_2013b,Tiwari_2014,Lu_2015,Lu_2015b}. 
Here, we study the interplay between topological order and superconducting correlations performing a symmetry analysis of the induced pair potential based on the anomalous Green function. We go beyond previous works \cite{Yokoyama_2012,Black-Schaffer_2012,*Black-Schaffer_2013,*Black-Schaffer_2013b} by including the effect of the magnetization on the superconducting electrode \cite{Tkachov_2015}. 

In the absence of magnetization, singlet and triplet components of the pair potential with even and odd frequency dependences are created at the TI-superconductor interface. However, the even-frequency singlet component is dominant and no low-energy states are formed \cite{Black-Schaffer_2012,*Black-Schaffer_2013,*Black-Schaffer_2013b}. 
For finite magnetization, the odd-frequency triplet components can become dominant, resulting in a rich subgap structure with emerging low-energy peaks. 
We study their effect on the NIS conductance, with I an intermediate region, and the local density of states (LDOS) for two orientations of the magnetization: in- and out-of-plane. 
For the latter, when the strength of the magnetization approaches the magnitude of the chemical potential in the superconductor, the odd-frequency triplet component becomes stronger than the singlet one. Consequently, a zero-energy peak emerges in the NIS conductance. 
We find the same behavior for in-plane case with strength comparable to the superconducting gap and oriented parallel to the NIS interface. 
When the in-plane magnetization is oriented differently, the rotational invariance of the band structure is broken. As a consequence, the NIS conductance becomes asymmetric with the energy; a feature completely different to previous works based on ferromagnet-superconductor junctions on 3DTIs \cite{Tanaka_2009,*Linder_2010b,Linder_2010,Lu_2015}. 
Independently of the orientation of the in-plane magnetization, when its strength is comparable to the superconducting gap, the odd-frequency triplet component becomes similar to or bigger than the singlet one. Consequently, subgap resonances are present in the LDOS and NIS conductance. 

This paper is organized as follows. 
We present the model for the TI-superconductor hybrid system and describe the effect of the magnetization in \sref{sec:model}.  
In \sref{sec:sym}, we describe the symmetry classification of the induced pair potential and study the conditions for the emergence of new symmetry terms for finite magnetization, which we associate with changes in the local density of states. 
In \sref{sec:cond}, we connect the previous results with an experimentally relevant observable like the NIS conductance. 
We present our conclusions in \sref{sec:conc}. 
Finally, in the Appendix, we describe our method to compute the retarded Green function and the transport observables that we examine. 


\section{Model \label{sec:model}}
In our setup, the surface of a 3DTI extends along the $x$-$y$ plane. The superconducting and intermediate regions are located, respectively, at $x>d$ and $0<x<d$ (see \fref{fig:sketch}). The role of the intermediate region is to include any interface scattering between the normal and the superconducting regions. We thus assume perfectly flat and clean interfaces. 
The low-energy electron and hole excitations at the surface of the 3DTI are described by the Bogoliubov-de Gennes (BdG) equations $\check{H}\Psi=E\Psi$, with $E$ the excitation energy. 
Particle-hole symmetry imposes that if $\Psi_E$ is a solution of the BdG equations with excitation energy $E$, $\Psi_{-E}$ must also be a solution. 
In Nambu (particle-hole) and spin space, with basis $\Psi=[\hat{c}_{\up}(\mbf{k}),\hat{c}_{\dw}(\mbf{k}),\hat{c}^{\dagger}_{\up}(-\mbf{k}),\hat{c}^{\dagger}_{\dw}(-\mbf{k})]^T$, with $\hat{c}_{\sigma}(\mbf{k})$ the annihilation operator for an electron of spin $\sigma=\up,\dw$ and momentum $\mbf{k}$, the Hamiltonian reads as
\begin{equation}
 \check{H}=\left(\!\begin{array}{cc}
                       \hat{h}(\mathbf{k})-\mu(x)\si & i\sy\Delta(x) \\ -i\sy\Delta(x) & \mu(x)\si-\hat{h}^{*}(-\mathbf{k})
                      \end{array}\!\right) \quad ,
\label{eq:hamil}
\end{equation}
where $\mu(x)$ is the chemical potential and the Pauli matrices $\hat{\sigma}_{0,1,2,3}$ act on spin space. 
We model the intermediate region as a square potential with thickness $d$ and height $\mu_I$, as it is shown in \fref{fig:sketch}(a). Therefore, we have
\begin{equation*}
 \mu(x)= \left\{ \begin{array}{cl} 
  \mu_N &, x\leq0 \\ 
  \mu_I &, 0<x<d \\ 
  \mu_S &, x\geq d 
 \end{array}\right. \quad .
\end{equation*}
We note that, due to Klein tunneling, the barrier does not affect the normally incident modes (with $k_y\sim0$) \cite{Tkachov_2012,Tkachov_2013}. 
The pair potential in the superconducting region is $\Delta(x)=\Delta\Theta(x-d)$, with $\Delta\ge0$ and $\Theta(x)$ the Heaviside function. Electron-like quasiparticles are described by a Dirac Hamiltonian as
\begin{equation}\label{eq:dirac}
 \hat{h}(\mathbf{k})= v_F\left(k_x\sx+k_y\sy \right)-e\mbf{A}\cdot\mbf{\hat{\sigma}} \quad ,
\end{equation}
with $v_F$ the Fermi velocity, $e$ the electron charge, and $\mbf{A}$ the vector potential (we take $\hbar=1$ and $c=1$ for simplicity). 
For a constant magnetic field, the vector potential depends linearly on the spatial coordinates. 
In what follows, we describe two situations where we can neglect the spatial dependence of $\mbf{A}$ and regard it as constant. 
The effect of a uniform constant $\mbf{A}$ has the same form as a Zeeman interaction where the vector potential plays the role of an exchange field. Consequently, the Hamiltonian in \eref{eq:hamil} is equivalent to the one used in, e.g., Ref.~\onlinecite{Tanaka_2009}, when we define an \textit{effective magnetization}
\begin{align}\label{eq:magnet}
 \check{M}={}&\left(\!\begin{array}{cc}
                       \mbf{m}\cdot\mbf{\hat{\sigma}} & 0 \\ 0 & -\mbf{m}\cdot\mbf{\hat{\sigma}}^{*}
                      \end{array}\!\right) \\ 
  ={}&-e\left[ \left(A_x\sx+A_z\sz\right)\otimes\tz +A_y\sy\otimes\ti\right]\quad , \nonumber
\end{align}
where the Pauli matrices $\tau_{0,1,2,3}$ act in Nambu space. 
The main difference with previous works \cite{Tanaka_2009,Linder_2010b,Snelder_2013} is that magnetic order is induced in the whole junction and is not just limited to an intermediate region. 
We consider this a sensible approach to describe 3DTI hybrid junctions where proximity-induced superconductivity on the two-dimensional surface state can be sensitive to magnetic effects. 

With our choice of the reference frame (see \fref{fig:sketch}), the momentum component parallel to the interface $k_y$ is conserved. Therefore, the solutions of the BdG equations are expressed as $\Psi(x,y)=\e^{ik_yy}\Psi(x,\theta)$, with $\sin\theta=k_y/|\mbf{k}|$. A full list of the scattering states for this junction is presented in Appendix \ref{sec:app1}. Next, we provide the main results for two different orientations of the effective magnetization $\check{M}$. 

\subsection{Out-of-plane effective magnetization}
We start by considering an out-of-plane effective magnetization $\check{M}=m_z\sz\otimes\tz$ (see \fref{fig:sketch}), 
stemming from a magnetic field such that $\mbf{A}=[0,0,A_z(x)]$. Ignoring the spatial dependence of the vector potential is equivalent to assume that the magnetization is a Zeeman term with exchange field $m_z$. 

In the superconducting region, the positive branch of the low-energy spectrum is then given by
\begin{equation}\label{eq:mz-energy} 
 E_{\pm}(\mbf{k}) = \sqrt{ \mu^2+\Delta^2 +v_F^2|\mathbf{k}|^2 + m_z^2 \pm 2 r_Z} \quad ,
\end{equation}
with $r_Z=\sqrt{\mu^2v_F^2|\mathbf{k}|^2+m_z^2 \left( \mu^2 + \Delta^2\right)}$. 
Although the energy spectrum looks rather complicated, its analysis is very simple if we focus on the $\Gamma$ point $\mbf{k}=0$, where we find
\[
 E_{\pm}(0) = |m_z \pm \sqrt{\mu^2+\Delta^2}| \quad .
\]
It is straightforward to see that both $m_z$ and the superconducting gap $\Delta$ independently open a gap in an otherwise gapless Dirac spectrum. When both $m_z$ and $\Delta$ are finite at the same time, a competition between the superconducting and the magnetic gap occurs. As a result, a zero-energy state can be created when $m_z^2=\mu^2+\Delta^2$. This allows us to define the effective superconducting gap
\begin{equation}
 \Delta_z\equiv\Delta\sqrt{1-(m_z/|\mu|)^2} \quad .
\label{eq:gaps_mz}
\end{equation}
For $0<m_z\le\mu$, the gap is real and superconductivity is the dominant process in the superconducting region. On the other hand, superconductivity is suppressed by a magnetic gap when $m_z>\mu>\Delta$. 
The critical point between the two regimes is thus $m_z=\mu>\Delta$. The Fermi energy for 3DTIs based on Bi compounds or strained HgTe can be estimated as $\mu\gtrsim50$ meV. A comparable exchange field can be obtained by proximity from a ferromagnetic insulator, as sketched in \fref{fig:sketch}(b). In a recent experiment, EuS was deposited on top of the topological insulator Bi$_2$Se$_3$. Ferromagnetic order induced onto the surface of the TI was estimated to be around $(60-400)$ meV/area [nm$^2$] per applied Tesla \cite{Moodera_2013}. For small enough devices, this correction can be much larger than the induced superconducting gap on TIs for fields below the critical field of the superconductor. 

\subsection{In-plane effective magnetization}
We now examine \eref{eq:magnet} in the heavily-doped, weak-field approximation with $\mu\gg |E|,\Delta,|\mbf{m}|$. Since the out-of-plane component becomes relevant only when $m_z\sim\mu$, we can ignore it in this approximation. The resulting in-plane effective magnetization is $\hat{M}=m_x\sx\tz+m_y\sy\ti\equiv m_{\parallel}\left(\sx\tz\cos\alpha+\sy\ti\sin\alpha\right)$, with the angle $\alpha$ measured from the $k_x$ direction [see more details in \fref{fig:sketch}(c)]. 
The in-plane effective magnetization is defined from the vector potential $\mbf{A}=[A_x(z),A_y(z),0]$ which we approximate by its value at the junction's plane. This approximation is also valid for the superconducting region if we let the penetration length of the superconductor be much larger than the superconducting coherence length \cite{Tkachov_2015,Sauls_1997,*Tanaka_2002,*Tanaka_2009b}. 

In the superconducting region, the positive branch of the low-energy spectrum is given by
\begin{equation}
 E_{\pm}(\mbf{k})\approx\mbf{n}\cdot\mbf{m} \!+\!\sqrt{\left(\mu\pm v_F|\mbf{k}|\right)^2+\Delta^2} \quad ,
\label{eq:hdwf_energy}
\end{equation}
where we have defined $\mbf{n}=\mbf{k}/|\mbf{k}|=(\cos\theta,\sin\theta,0)$. The effect of the weak field is a shift in the energy, analogous to the Doppler shift discussed for a SQUID-like geometry of helical edge states in Ref.~\onlinecite{Tkachov_2015}. 
Using that
\[
 \mbf{n}\cdot\mbf{m} = m_{\parallel}\left(\cos\theta\cos\alpha + \sin\theta\sin\alpha\right)= m_{\parallel} \cos\left(\theta-\alpha\right) \quad ,
\]
we can define the effective gaps
\begin{equation}
 \Delta_{\pm}\equiv |\Delta\pm m_{\parallel} \cos\left(\theta-\alpha\right)| \quad .
\label{eq:gaps_mxy}
\end{equation}
The critical value for which the effective gap closes is now $m_{\parallel}\!\sim\!\Delta$, which can be reached with rather weak fields. 
The magnetic field induces a finite Cooper pair (condensate) momentum which affects quasiparticles differently depending on their direction of motion. 
Due to the spin-momentum locking at the surface state of the TI, quasiparticles moving in opposite directions feel reversed magnetic fields. In other words, they travel upstream or downstream with respect to the superconducting condensate. This asymmetric behavior with the direction of motion is not found in non-topological systems \cite{Sauls_1997,*Tanaka_2002,*Tanaka_2009b,Burset_2009}. 

\subsection{Retarded Green function and experimental observables}
We now construct the retarded Green function associated to the Hamiltonian in \eref{eq:hamil}. The retarded Green function is obtained combining all scattering states as \cite{McMillan_1968,*Furusaki_1991,*Tanaka_1996,*Kashiwaya_2000,*Herrera_2010}
\begin{gather}\label{eq:GF-gen}
 \check{G}^r(x,x',E+i0^+;\theta)= \\
 \left\{\!\! \begin{array}{lr}
   \begin{array}{c} 
               \alpha_1\Psi_3(x)\tilde{\Psi}_1^T(x') \!+\! 				\alpha_2\Psi_3(x)\tilde{\Psi}_2^T(x') \\
               + \alpha_3\Psi_4(x)\tilde{\Psi}_1^T(x') \!+\! \alpha_4\Psi_4(x)\tilde{\Psi}_2^T(x')
   \end{array} \, , & x< x'
\\ 
   \begin{array}{c} 
               \beta_1\Psi_1(x)\tilde{\Psi}_3^T(x') \!+\! 				\beta_2\Psi_2(x)\tilde{\Psi}_3^T(x') \\
               + \beta_3\Psi_1(x)\tilde{\Psi}_4^T(x') \!+\! \beta_4\Psi_2(x)\tilde{\Psi}_4^T(x')
   \end{array} \, , & x> x'
               \end{array} \right. \quad . \nonumber
\end{gather}

An incoming electron (hole) from the normal region onto the superconductor can be 
\begin{inparaenum}[(\itshape a\upshape)]
\item Andreev reflected as a hole (electron); 
\item normal reflected as an electron (hole); 
\item transmitted to the superconductor as an electronlike (holelike) quasiparticle; and
\item as a holelike (electronlike) quasiparticle. 
\end{inparaenum}
We label this scattering process with the number $1$ ($2$). The corresponding wave function is thus $\Psi_{1(2)}(x)$. 
Analogously, the equivalent process where the incoming particle is an electronlike (holelike) excitation incident from the superconductor is labeled with the number $3$ ($4$). 
In Appendix \ref{sec:app2}, we show more details of the calculation of the Green function, the reflection amplitudes $a_j$, $b_j$, and the transmission amplitudes $c_j$ and $d_j$, with $j=1,\dots,4$. 
The wave functions $\tilde{\Psi}_i(x)$ correspond to the conjugate scattering processes to $\Psi_i(x)$. 
The coefficients $\alpha_i$ and $\beta_i$ are determined by the continuity equation obtained after integrating the BdG equations around $x=x'$, namely, 
\begin{gather}\label{eq:GF-cont}
 \frac{-i}{v_F}\sx\otimes \ti = \\ \check{G}^r(x,x+0^+,E+i0^+;\theta)-\check{G}^r(x,x-0^+,E+i0^+;\theta) \quad . \nonumber
\end{gather}

The spectral density of states is then calculated from the retarded Green function as
\begin{equation*}
 A(x,E;\theta)=-\frac{1}{\pi}\imag\left\{\mathrm{Tr}\check{G}_{ee}^r(x,x,E+i0^+;\theta)\right\} \quad ,
\end{equation*}
where the trace is taken on the electron-electron component of the Green function in Nambu space (i.e., the single-particle Green function). Furthermore, the local density of states (LDOS) is given by
\begin{equation}
 \rho(x,E)=\int\limits_{-\pi/2}^{\pi/2}\!\!\! \mathrm{d}\theta\cos\theta A(x,E;\theta) \quad .
\label{eq:ldos-def}
\end{equation}
In the next sections, we normalize the LDOS in the superconductor by the LDOS calculated deep inside the normal region, $\rho_N(E)=|\mu_N+E|/v_F$ (see Appendix \ref{sec:app3} for more details). 

We define the conductance of the NIS junction as \cite{BTK,Tanaka_1995}
\begin{equation}
 G_{NS}(E)=\int\limits_{-\pi/2}^{\pi/2}\!\!\! \mathrm{d}\theta\cos\theta \left[ 1-|b_1(E,\theta)|^2+|a_1(E,\theta)|^2\right] \quad ,
\label{eq:cond_NS}
\end{equation}
where $a_1(E,\theta)$ and $b_1(E,\theta)$ are the Andreev and normal reflection amplitudes, respectively, for an incident electron. 
We normalize the conductance using the constant value $G_0=G_{NS}(E\gg\Delta)$. 

Notice that both the conductance and the LDOS are averaged over all incident angles, labeled by $\theta\in[-\pi/2,\pi/2]$ \footnotemark[1]. 

\footnotetext[1]{The one-dimensional edge of a 2DTI, which is equivalent to the $\theta=0$ case considered here, presents a well-defined spin polarization. For the 2D surface state, the modes with $\theta\neq0$ usually spoil this neat effect. }


\section{Symmetry of the induced pair potential \label{sec:sym}}
Induced superconducting correlations are manifested in the anomalous Green function (the electron-hole component of the Green function in Nambu space), i.e., the $2\times2$ matrix in spin space
\[
 \tilde{F}_{\sigma\sigma'}(t_1,\mbf{r}_1;t_2,\mbf{r}_2)=\left\langle T_t\Psi_{\sigma}(t_1,\mbf{r}_1)\Psi_{\sigma'}(t_2,\mbf{r}_2) \right\rangle \quad ,
\]
with $T_t$ the time-ordering operator and $\Psi_{\sigma}(t_1,\mbf{r}_1)$ the fermion operator for given spin, time, and space coordinates. In homogeneous systems, we can define the relative coordinates $\mbf{r}=\mbf{r}_1-\mbf{r}_2$ and $t=t_1-t_2$ and the Pauli principle imposes that $\tilde{F}_{\sigma\sigma'}(t,\mbf{r})=-\tilde{F}_{\sigma'\sigma}(-t,-\mbf{r})$. After Fourier transformations, we have
\[
 F_{\sigma\sigma'}(\omega,\mbf{k})=-F_{\sigma'\sigma}(-\omega,-\mbf{k}) \quad .
\]
We can write explicitly the spin-singlet and triplet components of the anomalous Green function in the basis of \eref{eq:hamil} as
\[
 \hat{F}(\omega,\mbf{k})= \left[ F_0(\omega,\mbf{k})\si + F_j(\omega,\mbf{k})\hat{\sigma}_j\right]i\sy \quad .
\]
Evidently, the Pauli principle must be fulfilled independently by the singlet and the three triplet components. 
With respect to the spin degree of freedom, the singlet component is odd and the triplet ones even. Under a sign change in either the frequency or the momentum, the functions $F_\nu(\omega,\mbf{k})$, with $\nu=0,1,2,3$, can also be classified as even or odd. As a consequence, only four types of symmetries are allowed, namely, \begin{inparaenum}[(\itshape i\upshape)]
\item \textbf{E}ven-frequency spin-\textbf{S}inglet \textbf{E}ven-parity (ESE); 
\item \textbf{E}ven-frequency spin-\textbf{T}riplet \textbf{O}dd-parity (ETO); 
\item \textbf{O}dd-frequency spin-\textbf{S}inglet \textbf{O}dd-parity (OSO); and
\item \textbf{O}dd-frequency spin-\textbf{T}riplet \textbf{E}ven-parity (OTE) 
\end{inparaenum} [see \fref{fig:sketch}(b) and, for more details, Ref.~\onlinecite{Tanaka_2007,*Eschrig_2007,*Tanaka_2007b,*Tanaka_2007c,*Tanaka_JPSJ}]. 

In our model, we assume perfectly flat and clean interfaces which conserve the parallel component of the momentum ($k_y$ for the axis choice in \fref{fig:sketch}). In the quasi-one dimensional model, transport along these junctions is described by one-dimensional Green functions $\check{G}(x,x',\omega;k_y)$ separately for each channel $k_y$. 
For positive frequencies, we construct these Green functions using \eref{eq:GF-gen}, which considers scattering solutions of the BdG equations with excitation energies $E\ge0$. For negative frequencies, we need to consider the time-reversal of the scattering processes, i.e., the $E<0$ solutions of the BdG equations, which are provided by the advanced Green function. 
Therefore, to study the frequency dependence of the anomalous Green function, we combine the retarded and the advanced Green functions as 
\begin{gather}\label{eq:anom-GF}
 F_{\nu}[E+i\sgn(E)0^+] = \\ \Theta(-E)F^a_{\nu}(E-i0^+)+\Theta(E)F^r_{\nu}(E+i0^+) \quad , \nonumber
\end{gather}
where $E$ is a real variable, that coincides with the excitation energy for positive values, which we associate with the frequency. $k_y$ is parametrized using the angle of incidence $\theta$ and the advanced Green function is obtained using $\check{G}^a(x,x',E-i0^+;k_y)=[\check{G}^r(x',x,E+i0^+;k_y)]^{\dagger}$. 

\subsection{Zero-field analysis}
We now analyze each component of the anomalous Green function. 
We are mainly interested in local observables, like the LDOS which is determined by the spatial variation of the symmetry of the induced pair potential. Therefore, we only consider the case with $x=x'$ here. 
In Appendix \ref{sec:app2}, we show that the Green function is composed of an \textit{edge} and a \textit{bulk} part, i.e., one term given by the scattering at the interface and another by the solutions far away from the interface, respectively. For each component of the anomalous Green function we thus have
\begin{equation*}
 F_{\nu}(E,\theta)= F_{\nu,\infty}(E,\theta)+ F_{\nu,A}(E,\theta) + F_{\nu,B}(E,\theta) \quad .
\end{equation*}
$F_{\nu,\infty}$ corresponds to the Green function at $x\!=\!x'\!\rightarrow\!\infty$, i.e., the \textit{bulk} term inside the superconducting region. On the other hand, $F_{\nu,A}$ and $F_{\nu,B}$ are the \textit{edge} Green functions proportional to Andreev and normal reflection processes, respectively. In what follows, we ignore $F_{\nu,B}$ since it only includes rapid oscillations. 

In the absence of magnetization, for the bulk terms we immediately see that 
\begin{align*}
 F_{0,\infty}(E,\theta)={}& \frac{i\sgn\left(E\right)}{2v_F|\cos\theta|} \frac{\Delta} {\sqrt{E^2-\Delta^2}} \quad , \\
 F_{2,\infty}(E,\theta)={}& \frac{i\sgn\left(E\right)}{2v_F|\cos\theta|} \frac{\Delta} {\sqrt{E^2-\Delta^2}}\sin\theta \quad , \\
 F_{1,\infty}(E,\theta)={}& F^r_{3,\infty}(E,\theta) = 0 \quad ,
\end{align*}
where 
\begin{gather*}
 \sqrt{E^2-\Delta^2} = \\ \left\{ \begin{array}{cr}
               \sgn\left(E\right)\sqrt{[E\!+\!i\sgn\left(E\right)0^{+}]^2-\Delta^2} & , \, |E|>\Delta \\ i\sgn\left(E\right)\sqrt{\Delta^2-[E\!+\!i\sgn\left(E\right)0^{+}]^2} & , \, |E|\le\Delta  \end{array} \right. \quad ,
\end{gather*}
which is an odd-function of the energy. Therefore, we find that $F_{0,\infty}$ and $F_{2,\infty}$ are even-frequency functions; they are thus classified as ESE and ETO terms, respectively. Even in the absence of magnetic order, the combination of a spin-orbit induced Dirac spectrum with conventional spin-singlet superconductivity results in a superposition of singlet and triplet superconducting correlations \cite{Read_2000,Gorkov_2001,Schnyder_2008,Yokoyama_2012,Tkachov_2012,Tkachov_2013,Tkachov_2013b,Burset_2014}. 
The ETO term is proportional to $\sin\theta=k_y/|k|$ because it is odd in momentum. Therefore, after averaging over incident angles, any odd-momentum terms must vanish. 
Consequently, the effect of ETO terms on a local observable in our system is canceled by the average over incident angles. 

At the IS interface with $x=d$, the breaking of spatial inversion generates odd-frequency terms both for the singlet and triplet components \cite{Black-Schaffer_2012,*Black-Schaffer_2013,*Black-Schaffer_2013b,Tanaka_2007,*Eschrig_2007,*Tanaka_2007b,*Tanaka_2007c,*Tanaka_JPSJ}. Angle-averaging cancels any odd-frequency singlet term, but the odd-frequency triplet components remain. 
We thus find
\[
 F_{1,A}(E,\theta)=\frac{i}{v_F}a_3(E,\theta)\e^{i\frac{2\sqrt{E^2-\Delta^2}}{v_F\cos\theta} x} \quad ,
\]
with $a_3(E,\theta)$ the Andreev reflection amplitude for an electron-like quasiparticle incident from the superconductor, which results in an OTE contribution that survives the integration over incident angles. 

To compare the relative weight of the triplet components against the singlet one, we define the magnitude of the angle-averaged triplet vector 
\begin{equation}\label{eq:triplet}
 F_t(E)=\left[ \sum\limits_{j=1}^{3} \left| \int\limits_{-\pi/2}^{\pi/2}\!\!\!\mathrm{d}\theta \cos\theta F_{j}(E,\theta) \right|^2 \right]^{1/2}\quad .
\end{equation}
Due to the integration over incident angles, $F_t(E)$ represents only odd-frequency triplet terms. Therefore, at zero field, $F_t(E)=|F_{1,A}(E)|$. It is interesting to note that, in our spin basis, this term is given by $F_1\propto \left(\up\up-\dw\dw\right)$, while the opposite polarization, $F_2\propto \left(\up\up+\dw\dw\right)$, vanishes. This is a consequence of our calculation being limited to one surface of the 3DTI; for the other surface the polarization of the Cooper pairs at the interface is reversed. 


\begin{figure}
	\includegraphics[width=\columnwidth]{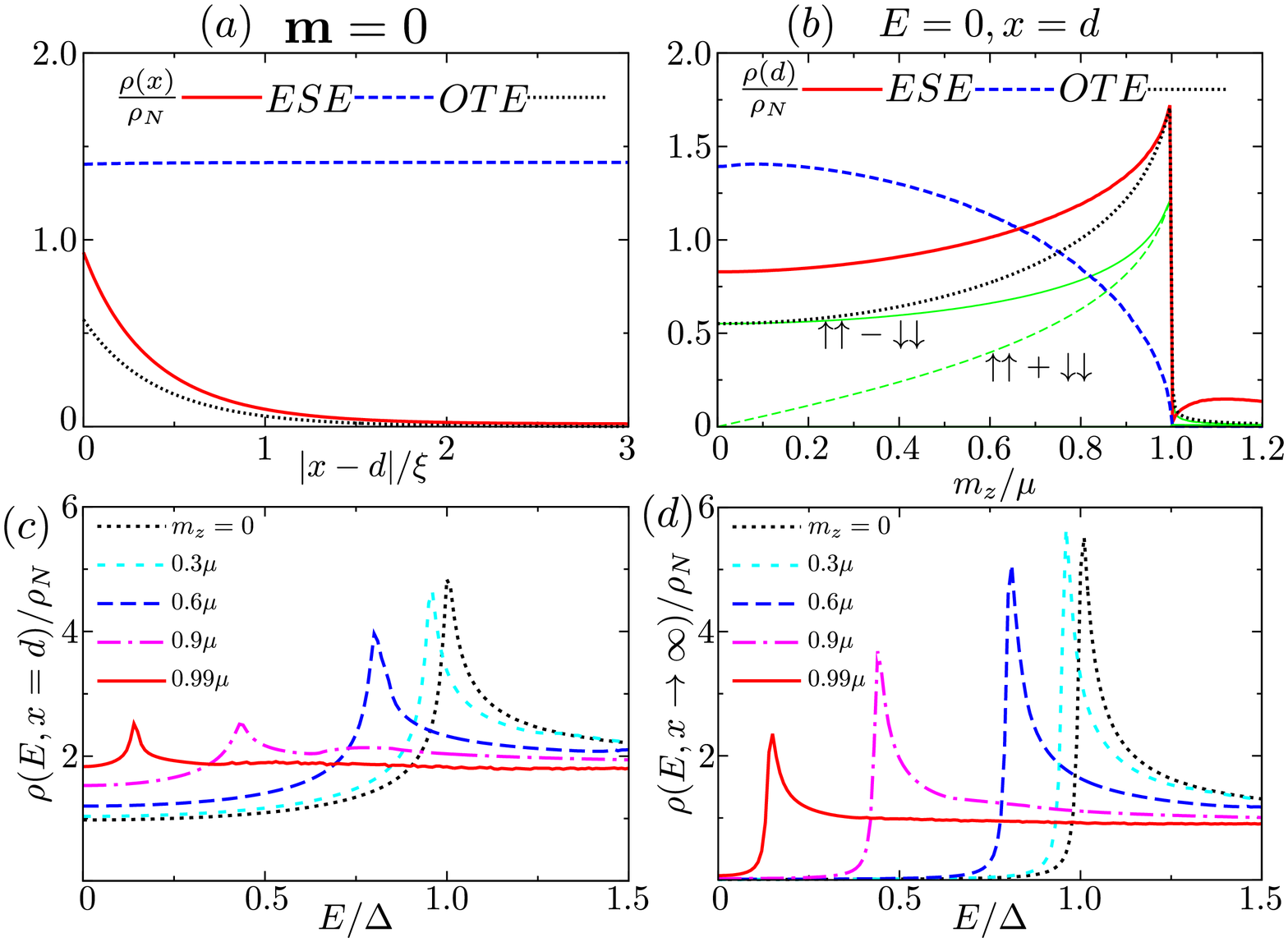}
	\caption{\label{fig:fdos_z}
	 (a) Evolution of $\rho(E=0)$, angle-averaged $|F_{0}|$, and $F_t$ with the distance inside the superconducting region from the interface at $x=d$ at zero magnetization and $E=0$. (b) Same quantities as before as a function of  $m_z$, for $E=0$ and $x=d$. The green solid and dashed lines correspond to the two triplet polarizations $|F_{1,2}|\propto(\up\up\mp\dw\dw)$, respectively. (c,d) LDOS as a function of the energy for several values of $m_z$ calculated at the IS interface $x=d$ (c), and inside the superconducting region, where $x\rightarrow\infty$ (d). For all plots, we set $\mu_N=10\mu$, $\mu_I=20\mu$, $\mu_S=\mu=10^3\Delta$, and $d=0.1(v_F/\mu)$. }
\end{figure}


We show in \fref{fig:fdos_z}(a) the evolution of the angle-averaged quantities $|F_{0}(E=0)|$ and $F_t(E=0)$ as a function of the distance inside the superconducting region (dashed blue and dotted black lines, respectively). The bulk ESE contribution is dominant and responsible for the suppression of the density of states (red solid line) for $x>\xi$. At the interface, however, we find that $F_{t}(E=0)\neq0$. The OTE contribution vanishes inside the superconducting region but is responsible for the enhancement of the LDOS close to the interface. 

\subsection{Symmetry analysis for out-of-plane effective magnetization}
We now study the local anomalous Green function for an out-of-plane effective magnetization. We present a list of the components of the anomalous Green function in Appendix \ref{sec:app4}, for the perpendicular case treated here and for the in-plane one analyzed in the next section. The presence of the perpendicular magnetization affects all components and, in particular, induces two new bulk terms. $F_{1,\infty}$ is proportional to $k_y=|k|\sin\theta$ (ETO) and becomes zero after integration over $\theta$. On the other hand, $F_{3,\infty}$ is an OTE term proportional to $E m_z$ and it vanishes in the absence of magnetization or for $E=0$. Therefore, the main effect on the bulk region is determined by $F_0$. 

The situation is different at the interface. We plot in \fref{fig:fdos_z}(b) angle-averaged $|F_{0}(E=0)|$ and $F_t(E=0)$ (blue dashed and black dotted lines, respectively), as a function of $m_z$, for $x=d$. We choose $E=0$ to analyze the symmetry of any zero-energy peak in the LDOS (red solid line). 
The \textit{edge} contributions of the triplet components have odd-frequency parts that remain finite after integrating over the incident angles. $F_t(0)$ becomes comparable to $|F_{0}(0)|$ for a magnetization strength $m_z\approx0.7\mu$. 
Additionally, we plot the triplet components $|F_{1,2}(E=0)|\propto(\up\up\mp\dw\dw)$ using solid and dashed green lines, respectively. At $m_z=0$, only $|F_1(0)|$ is present. For a finite $m_z$, $|F_2(0)|$ becomes a finite OTE term and increases with the field strength like $|F_1|$ at the same rate that the ESE term is suppressed. The LDOS at the interface is enhanced accordingly. 
With the increase of $m_z$, the feature of a Majorana resonant state becomes more pronounced. Also, here Majorana resonances are always accompanied by odd-frequency pairing. 

The magnetization reduces the bulk gap $\Delta_z$, given in \eref{eq:gaps_mz}, for $m_z\leq\mu$. After the gap closes, the LDOS and the anomalous terms of the Green function are strongly suppressed. 
We show the energy dependence of the LDOS at the interface with $x=d$ in \fref{fig:fdos_z}(c) and in the bulk of the superconducting region with $x\rightarrow\infty$ in \fref{fig:fdos_z}(d). 
At the interface, the LDOS features a strong resonance at $|E|=\Delta_z$ and is finite for $|E|<\Delta_z$. 
For $m_z>\mu$, the LDOS is still finite and maximum at $E=0$ [see \fref{fig:fdos_z}(b)]. However, the value of this finite LDOS at the interface decreases rapidly when increasing $m_z$ over $\mu$ because it is suppressed by a magnetic gap. 
Deep inside the superconducting region, where only the ESE component of the anomalous Green function is finite, we find a BCS-type density of states around the effective gap $\Delta_z$. For $m_z>\mu$, a magnetic gap opens destroying superconductivity. 


\begin{figure}
	\includegraphics[width=\columnwidth]{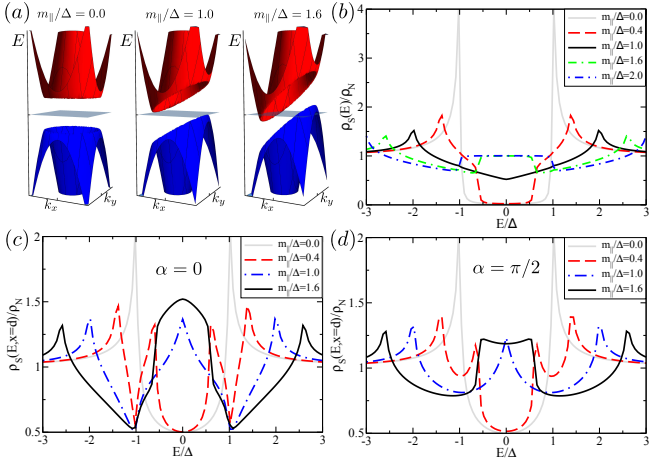}
	\caption{\label{fig:LDOS_mxy}
	(a) Energy-momentum plot of the low-energy bands inside the superconducting region for several values of $m_{\parallel}$. (b) Bulk LDOS of the superconducting region of the NIS junction as a function of the excitation energy $E$ for different $m_{\parallel}$. (c,d) LDOS at the interface $x=d$ with $\mu_I/\mu=10$ and $d=0.1(v_F/\mu)$ as a function of the excitation energy $E$ with different in-plane magnetizations. We show $\alpha/\pi=0,0.5$ for (c,d), respectively. 
	For (b,c,d) we set $\mu=10^{3}\Delta$. }
\end{figure}

\subsection{Symmetry analysis for in-plane effective magnetization}
In the previous section, we showed that the effect of a magnetic field in the heavily-doped, weak-field approximation is a Doppler-type shift of the excitation energy. This shift has an important effect on the energy bands: in \fref{fig:LDOS_mxy}(a), we plot the conduction (red) and valence (blue) bands in momentum space for several values of $m_{\parallel}$. For reference, the zero-energy plane is marked by a gray area. At zero field, the spectrum has rotational symmetry and the superconducting gap is the same for $|\mbf{k}|=k_F$. For finite fields, $m_{\parallel}\neq0$, the rotational symmetry of the spectrum is broken. However, particle-hole symmetry is recovered using $\epsilon(\mbf{k})=-\epsilon(-\mbf{k})$. The orientation of the NIS interfaces with respect to this bulk spectrum becomes very important since the energy spectrum can be asymmetric around $k_y=0$, with $k_y$ the momentum component parallel to the interface. Only the case where the in-plane field is oriented 
parallel to the normal-superconductor interface, i.e., for $\alpha=\pi/2$, the energy bands are $k_y$-symmetric. 
Additionally, there is a \textit{critical} magnetization $m_{\parallel}=\Delta$ where both bands touch the zero-energy plane and the superconducting gap closes at some values of $|\mbf{k}|=k_F$, for any orientation of the in-plane field. The orientation of the in-plane field $\alpha$ changes the position in momentum space where this closing occurs \cite{[{The gap closing can greatly affect the edge states of the two-dimensional TI, as recently described in }] [{}]Reinthaler_2015}. 

To study the effect of the gap closing on the bulk of the superconducting region, we show the bulk density of states in \fref{fig:LDOS_mxy}(b) for several values of $m_{\parallel}$. After averaging over the angle of incidence, the bulk density of states is independent of the orientation of the magnetization. When $0<m_{\parallel}<\Delta$, the effective gaps $\Delta_{\pm}$ defined in \eref{eq:gaps_mxy} emerge and the LDOS exhibits a peak at $E\sim|\Delta_+|$ and is gapped for $E\lesssim|\Delta_-|$. When the gap closes for $m_{\parallel}=\Delta$, the bulk density of states becomes V-shaped with two resonances at $E=\pm2\Delta$. 
For this case, the LDOS at the interface features a sharp zero-energy peak for any orientation of the in-plane field, as it is shown in the dashed-dotted blue lines of \fref{fig:LDOS_mxy}(c,d). 
Beyond this point, the zero-energy peak becomes wider if $\alpha<\pi/4$ or splits into two otherwise. For the bulk LDOS, a region emerges around $E=0$ with a flat density of states of the same magnitude as that of the normal region. Outside this energy range, the V-shaped profile becomes wider with increasing the strength of the field. 


\begin{figure}
	\includegraphics[width=\columnwidth]{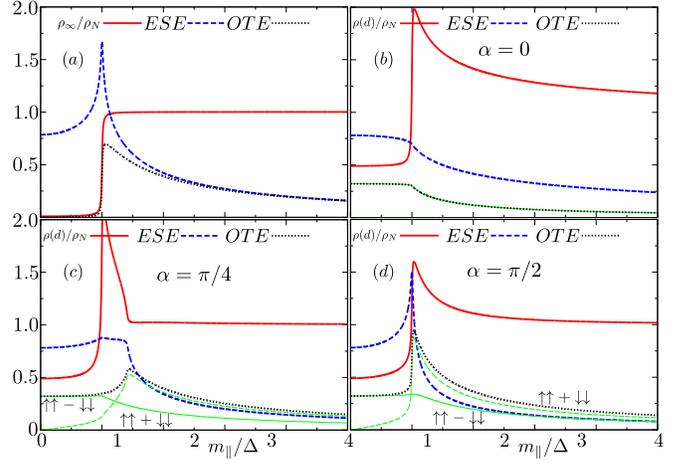}
	\caption{\label{fig:fdos}
	 Plots of $\rho(E=0)$, $|F_{0}(E=0)|$, and $F_t(E=0)$ as a function of $m_{\parallel}$. (a) Bulk results for $x\rightarrow\infty$. (b,c,d) Results at the interface with $x=d$ for different orientations of the effective magnetization: $\alpha/\pi=0,0.25,0.5$ for (b,c,d), respectively. The solid and dashed green lines correspond to the polarizations $\up\up-\dw\dw$ and $\up\up+\dw\dw$ of the OTE term, respectively. For all plots, $\mu_N=\mu_S=\mu=10^{3}\Delta$, $\mu_I/\mu=10$, and $d=0.1(v_F/\mu)$. }
\end{figure}


We now study the components of the anomalous Green function plotting the bulk contribution of $|F_{0}(E=0)|$ and $F_t(0)$ as a function of $m_{\parallel}$ in \fref{fig:fdos}(a). There are two distinct regimes. For $m_{\parallel}<\Delta$, the ESE term is greatly enhanced, while the OTE is finite but small compared to it. The LDOS is suppressed which corresponds to a conventional gapped profile. 
For the critical magnetization $m_{\parallel}=\Delta$, all terms increase. 
At this point, the LDOS jumps from $0$ to $1$. From then on, the triplet vector magnitude is finite and similar to the ESE term for $m_{\parallel}\gtrsim 2\Delta$ while the LDOS remains constant. Therefore, a finite bulk LDOS is obtained because of an equal superposition of OTE and ESE terms in the bulk of the superconducting region. 

We show in \fref{fig:fdos}(b,c,d) the ESE and OTE components at the interface ($x=d$), as a function of $m_{\parallel}$ for different orientations of the in-plane magnetization [$\alpha=0$, $\pi/4$, and $\pi/2$ for Figs. 4(b), 4(c), and 4(d), respectively]. 
For the symmetric orientation $\alpha=\pi/2$, the OTE term becomes dominant over the singlet one for values equal or greater than the critical magnetization [\fref{fig:fdos}(d)]. 
In the opposite case with $\alpha=0$ [\fref{fig:fdos}(b)], the singlet term is always dominant over the triplet one. Independently of the orientation, the LDOS is greatly enhanced for values greater than the critical one. 

Finally, the behavior of $|F_{1,2}|\propto(\up\up\mp\dw\dw)$ strongly depends on the orientation of the magnetization (see green solid and dashed lines for $|F_{1,2}|$, respectively). Indeed, for $\alpha=0$ we find that $|F_{1}|=|F_{2}|$ while they are very different for $\alpha\neq0$. Therefore, the spin polarization of the induced Cooper pairs can be controlled by the orientation of the external field \footnotemark[2].  

\footnotetext[2]{The orientation of the in-plane vector potential $\mbf{A}$ and, hence, that of the effective magnetization $\mbf{m}_{\parallel}$, is shifted from that of the magnetic field by $-\pi/2$. }

\section{Conductance spectroscopy \label{sec:cond}}
Our setup, sketched in \fref{fig:sketch}, is ideal for conductance spectroscopy measurements like the ones recently performed in Refs. \onlinecite{Finck_2014,Brinkman_2015}. The conductance of the NIS junction, obtained from \eref{eq:cond_NS}, is normalized to $G_0=G_{NS}(E\gg\Delta)$, as it is commonly done in experiments. 


\begin{figure}
	\includegraphics[width=\columnwidth]{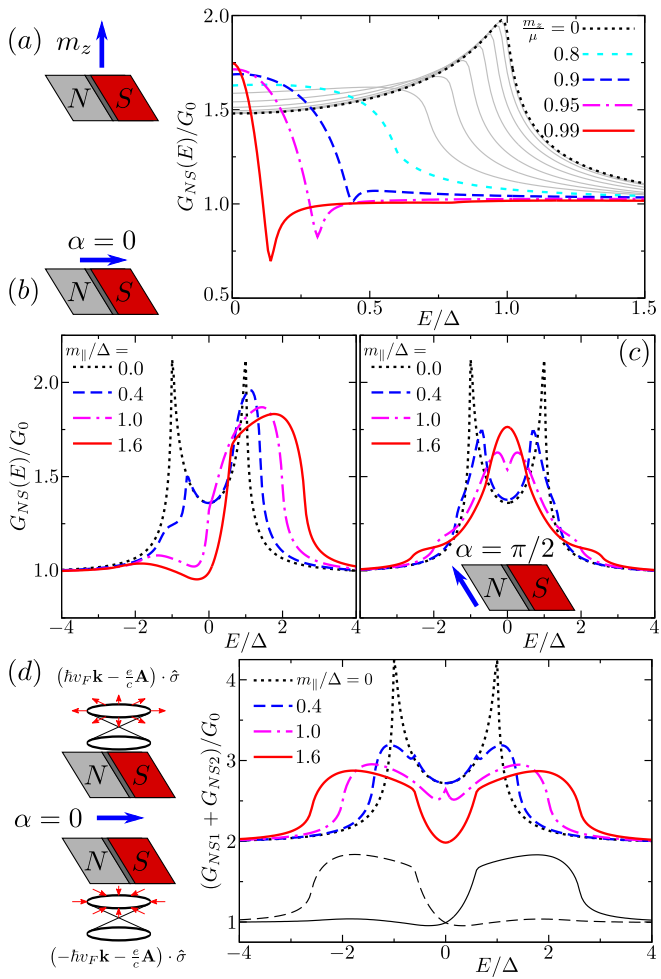}
	\caption{\label{fig:NScond}
	 Conductance as a function of the excitation energy for several effective magnetizations. For each case we include a sketch of the NIS junction displaying the orientation of the effective magnetization as a blue arrow. (a) A perpendicular magnetization is finite in the whole NIS junction with $\mu_N=10\mu$, $\mu_I=20\mu$, and $\mu_S=\mu$. The thin gray lines are taken at intervals $0.1\mu$ from $m_z=0$. (b,c) In-plane magnetization finite in the whole junction oriented perpendicular (b) or parallel (c) to the NIS interfaces. For both cases, $\mu_N=\mu_S=\mu$ and $\mu_I/\mu=10$. (d) With the same parameters as (b), superposition of the conductance for both surfaces; sketched on the left. The black solid line corresponds to the $m_{\parallel}/\Delta=1.6$ case of (b) while the dashed one is the conductance of the other surface with the same parameters. In all plots, $d=0.1(v_F/\mu)$ and $\mu=10^3\Delta$. }
\end{figure}


We start considering an out-of-plane effective magnetization finite in the whole junction. 
Consequently, when we approach the critical point where $m_z\sim\mu$, a magnetic gap starts to develop in both normal and superconducting regions. In order to have propagating states in the normal region, we choose $\mu_N=10\mu$ and $\mu_S=\mu=10^3\Delta$. We still include a square potential barrier setting $\mu_I=20\mu$. The results for this setup are shown in \fref{fig:NScond}(a) for several values of $m_z$ in the range $0<m_z<\mu$. 
For $m_z=0$, the tunnel conductance features a gapped profile. Due to the Klein tunneling effect, the conductance is not fully gapped for $E<\Delta$, even in the presence of a strong barrier. 
For $0<m_z<\mu$, the effective gap is reduced and, as $m_z$ approaches $\mu$, the subgap conductance increases clearly featuring a zero-bias peak. 
The formation of this zero-bias peak coincides with the regime where the odd-frequency triplet $F_t$ becomes dominant with respect to the ESE term $|F_0|$ [see \fref{fig:fdos_z}(b)]. For $m_z>\mu$, the gap in the excitation spectrum is given by $E_z=|m_z-\sqrt{\mu^2+\Delta^2}|$ and the conductance is zero for $|E|\leq E_z$. 

We now consider an in-plane magnetization in the heavily-doped, weak-field approximation. Under this approximation, if we take $\mu_N\sim\mu_S\gg |E|,\Delta,m_{\parallel}$, the effect of the field in the normal and intermediate regions is almost negligible. Assuming a setup with a symmetric square potential barrier with $\mu_N=\mu_S=\mu$ and $\mu_I=10\mu$, we plot in \fref{fig:NScond}(b,c) the conductance for several values of $m_{\parallel}$. 
We show an in-plane magnetization perpendicular to the NIS interfaces ($\alpha=0$) in \fref{fig:NScond}(b) and parallel to them ($\alpha=\pi/2$) in \fref{fig:NScond}(c). 
In both cases, the effect of the field is to split the superconducting gap into two effective gaps $\Delta_{\pm}$ defined in \eref{eq:gaps_mxy}. 

For $\alpha=\pi/2$, the resonance at $E=\Delta$ for $m_{\parallel}=0$ remains pinned to the smallest of the gaps, but the one at the biggest gap is smeared by the angle average. Due to Klein tunneling, these resonances can not be clearly resolved in the conductance, although they appear in the LDOS, as shown in \fref{fig:LDOS_mxy}(c,d). The smallest of the gaps closes for $m_{\parallel}\gtrsim\Delta$ and a zero-energy peak appears in the conductance. It is in this regime where the odd-frequency triplet term is greater than the singlet one, as shown in \fref{fig:fdos}(d). 

Strikingly, for $\alpha\neq\pi/2$, the conductance is asymmetric with the energy. 
This asymmetry is a consequence of the distortion of the energy bands by the magnetic field, as shown in \fref{fig:LDOS_mxy}(a). The field breaks the rotation symmetry of the energy bands and the modified bands have a point-like symmetry that is determined by the orientation of the in-plane magnetization. Since we study transport in the direction perpendicular to the NIS interfaces, when the magnetization is aligned with the interface, i.e., $\alpha=\pi/2$, the symmetry is recovered [see \fref{fig:NScond}(c)], while it is lost for any other orientation. 

Owing to the special particle-hole symmetry of our system, we find that, for $\alpha\neq\pi/2$, Andreev reflection is very asymmetric with the energy while normal reflection is symmetric. This effect holds independently of the barrier strength provided that the effective magnetization is misaligned with the NIS interface. Thus, as an example, we can consider a junction with $\mu_I=\mu_S\neq\mu_N$. For such junctions, normal and Andreev reflections take place separately at each side of the intermediate region. In the setup of \fref{fig:sketch}(a), when $|E|\le\Delta$, normal (Andreev) reflection is the only scattering process at $x=0$ ($x=d$). The reflection amplitudes are then
\begin{subequations}\label{eq:amplitudes}
\begin{align}
 a_1(E,\theta)={}&\frac{2\cos\theta_N\cos\theta_S}{t(E,\theta)}\Gamma_- \quad , \\
 b_1(E,\theta)={}&i\frac{\sin\theta_N-\sin\theta_S}{t(E,\theta)}\left(1-\Gamma_+\Gamma_-\right) \quad , \\
 t(E,\theta)={}& 1+\cos\left(\theta_N+\theta_S\right) \nonumber \\ & - \Gamma_+\Gamma_-\left[1-\cos\left(\theta_N - \theta_S\right)\right] \quad , \nonumber
\end{align}
\end{subequations}
with $\theta_{N,S}=\sin^{-1}(k_y/\mu_{N,S})$ and 
\begin{align*}
 \Gamma_{\pm}={}&\frac{\Delta^*}{\epsilon_{\pm}+\sqrt{(\epsilon_{\pm})^2-|\Delta|^2}}
 \quad , \\
 \epsilon_{\pm}={}& E\pm m_{\parallel}\cos(\theta_S-\alpha) \quad .
\end{align*}

Due to the spin-momentum locking at the surface of the 3DTI, the magnetic field affects differently quasiparticles moving in opposite directions \cite{Reinthaler_2015}. That is the reason why Andreev reflection is asymmetric with the magnetization, being proportional to $\Gamma_-$ while normal reflection is not, since it goes with $(1-\Gamma_+\Gamma_-)$. 
As a consequence, the conductance becomes more asymmetric as we increase $m_{\parallel}$, as shown in \fref{fig:NScond}(b). The energy range where Andreev reflection is enhanced also becomes wider. In the figure, this happens always for positive energies. 

Finally, the asymmetry in the conductance is also partly a consequence of having limited our analysis of transport to only one surface of the 3DTI. In \eref{eq:hamil}, the Dirac Hamiltonian chosen to describe electron-like quasiparticles is $\hat{h}(\mbf{k})=[v_F \mbf{k}-e\mbf{A}]\cdot\hat{\sigma}$. With this choice, quasi-particles have positive helicity, i.e., their spin points in the direction of their momentum. For the opposite surface state of the 3DTI, however, electron-like quasiparticles are described by $\hat{h}(\mbf{k})=[-v_F \mbf{k}-e\mbf{A}]\cdot\hat{\sigma}$. Therefore, the sign of their helicity changes and their spin points in the opposite direction of their momentum. As a consequence, the effect of the in-plane magnetization is reversed, with $\Gamma_{\pm}\rightarrow\Gamma_{\mp}$. The normal reflection amplitude is unchanged, but the Andreev reflection amplitude becomes proportional to $\Gamma_+$ in \eref{eq:amplitudes}. The asymmetry of the conductance is reversed, as it is shown by the black solid and dashed lines of \fref{fig:NScond}(d), corresponding to the $m_{\parallel}/\Delta=1.6$ case of \fref{fig:NScond}(b). If we consider a conductance measurement that includes opposite surfaces of the 3DTI, the symmetry with the energy is recovered. For the same parameters as \fref{fig:NScond}(b), we show such conductance spectroscopy in \fref{fig:NScond}(d). 


\section{Conclusions \label{sec:conc}}
We have analyzed the superconducting proximity effect at 3DTI-superconductor hybrids in the presence of an effective magnetization. We find that the interplay between the magnetization and the spin-momentum locking of the surface state gives rise to interesting odd-frequency triplet terms in the induced pair potential. Contrary to the usual even-frequency triplet terms, averaging over the angle of incidence does not cancel these components, and their effect can be observed in experimental transport observables such as the NIS conductance and LDOS. 

For an out-of-plane magnetization, the OTE terms increase with the strength of the field in the regime $\mu>m_z>0$. At the same time, the singlet term of the induced pair potential is suppressed. As a consequence, the conductance evolves from a conventional gapped profile for $m_z=0$ into a zero-bias conductance peak when $m_z\lesssim\mu$. 
Such strong magnetization can be induced in the system when a ferromagnetic insulator such as EuS is deposited on top of the NIS junction \cite{Moodera_2013}, in a setup like the one sketched in \fref{fig:sketch}(b). 

If we consider the effect of an orbital magnetic field on the junction, we find a similar gap to zero-energy peak transition on the conductance for in-plane magnetization parallel to the NIS interface. 
The main difference is that the critical value that determines the emergence of a zero-energy peak is now $m_{\parallel}=\Delta$, where the energy bands feature an indirect closing of the gap. 
For $m_{\parallel}\ge\Delta$, the OTE term is greater than the ESE one at the interface and similar in magnitude in the bulk superconducting region. As a result, we have connected the emergence of odd-frequency triplet in the induced pair potential on the surface of the 3DTI with a distinctive profile of the conductance. 

When the in-plane magnetization is not oriented parallel to the interface, the conductance becomes asymmetric with the energy. This is a manifestation of the breaking of the rotational symmetry of the band structure of the combined 3DTI-superconductor system. Only a transport measurement that probes one surface state of the 3DTI can detect the asymmetry of the conductance. Such setup, would be a perfect platform to probe the coexistence of topological order and superconductivity. 

Our results show that 3DTI-based NIS junctions feature a very rich induced superconducting pair potential that displays the elusive odd-frequency triplet component. 
Using very basic ingredients such as a conventional superconductor and an external magnetic field or ferromagnetic insulator, standard experimental techniques like conductance spectroscopy can be used to detect signatures of OTE superconductivity. 


\section*{Acknowledgments}
We thank D. Bercioux, F. S. Bergeret and F. Cr\'{e}pin for helpful discussions and comments. 
We acknowledge financial support by the DFG-JST research unit ``Topotronics'', the Helmholtz Foundation (VITI), the DFG Priority Program SPP1666, FOR1162, SFB 1170 ``ToCoTronics'', the ENB Graduate School on ``Topological Insulators'', and DFG Grant No. TK 60/1-1. 
This work was also supported by Topological Materials Science (TMS) (Grant No. 15H05853) and Grant No. 25287085 from the Ministry of Education, Culture, Sports, Science, and Technology, Japan (MEXT), and by the Core Research for Evolutional Science and Technology (CREST)
of the Japan Science and Technology Corporation (JST). 

\appendix

\section{Scattering states \label{sec:app1}}
In this appendix, we present the solutions of the BdG equations $\check{H}\Psi=E\Psi$, for the two orientations of the effective magnetization considered in the main text. 
In the normal regions with $\Delta=0$, the general solution of the BdG equations is 
\begin{align}
 \Psi(x<d)={}& A_{N}^{+}\psi_{+k_e}\e^{ik_e x} + A_{N}^{-}\psi_{-k_e}\e^{-ik_e x} \nonumber \\ +&  B_{N}^{+}\psi_{+k_h}\e^{ik_h x} + B_{N}^{-}\psi_{-k_h}\e^{-ik_h x} \quad , 
\label{eq:wvfunction_N}
\end{align} 
where the coefficients $A_N^{\pm}$ ($B_N^{\mp}$) represent the amplitudes for electrons (holes) propagating to the right and left, respectively. 

On the other hand, for the superconducting region we find 
\begin{align}
 \Psi(x>d)={}& A_{S}^{+}\psi_{+k_1}\e^{ik_1 x} + A_{S}^{-}\psi_{-k_1}\e^{-ik_1 x} \nonumber \\ +& B_{S}^{+}\psi_{+k_2}\e^{ik_2 x} + B_{S}^{-}\psi_{-k_2}\e^{-ik_2 x} \quad ,
\label{eq:wvfunction_S}
\end{align}
where the coefficients $A_S^{\pm}$ ($B_S^{\mp}$) now label the amplitudes for right and left moving electron-like (holes-like) quasiparticles. 

\subsection{Junctions with out-of-plane effective magnetization}
In the normal state regions with $\Delta=0$, the wave vector is given by $k_{e,h}= \sqrt{Z_{e+}Z_{e-}-(v_Fk_y)^2}/(v_F)$, with $Z_{e\pm}=\mu+E\pm m_z$ and $Z_{h\pm}=\mu-E\pm m_z$. The wave functions in \eref{eq:wvfunction_N} are then given by
\begin{subequations}\label{eq:wvf_N}
\begin{align}
 \psi_{\pm k_e}={}& \left[ 1, \pm\e^{\pm i\theta_{e}}, 0, 0 \right]^T \quad , \\
 \psi_{\pm k_h}={}& \left[ 0, 0, 1, \mp s_h \e^{\pm i\theta_{h}} \right]^T \quad ,
\end{align}
\end{subequations}
where we have defined the phase factors 
\begin{equation*}
 \e^{\pm i\theta_e}=v_F\frac{k_e \pm ik_y}{|Z_{e+}|} 
 \quad , \quad
 \e^{\pm i\theta_h}=v_F\frac{k_h \pm s_h ik_y}{|Z_{h-}|} 
 \quad ,
\end{equation*}
and $s_h=\sgn(\mu-E)$. 

Analogously, for the superconducting region, the wave vectors are given by $k_{j} = \sqrt{ Z_{j+}Z_{j-}-(v_Fk_y)^2 }/(v_F)$, with $j=1,2$,  
\begin{align*}
 Z_{1(2)\pm}={}&\frac{\mu^2-m_z^2 +(-)\mu \Omega_z \pm E m_z}{\mu \pm m_z} \quad , \\
 \Omega_z ={}& \left\{ \begin{array}{cl} \sgn\left(E\right)\sqrt{E^2-\Delta_z^2} & , |E|>\Delta_z \\ i\sqrt{\Delta_z^2-E^2} & , |E|\leq\Delta_z  \end{array} \right. \quad ,
\end{align*}
and $\Delta_z=\Delta\sqrt{1-(m_z/\mu)^2}$. 
The corresponding wave functions in \eref{eq:wvfunction_S} are
\begin{subequations}\label{eq:wvf_Sz}
\begin{align}
  \psi_{\pm k_1}={}& \left[\Gamma_1,\pm E_{1\pm}\Gamma_1,\mp \zeta E_{1\pm}, 1 \right]^T \quad , \\
  \psi_{\pm k_2}={}& \left[\Gamma_2,\pm E_{2\pm}\Gamma_2,\mp \zeta E_{2\pm}, 1 \right]^T \quad , 
\end{align}
\end{subequations}
with
\begin{align*}
  E_{1(2)\pm} ={}& \frac{Z_{1(2)+}}{v_F(k_{1(2)} \mp ik_y)} \quad , \\
  \Gamma_{1,2}={}& \frac{\mu E \pm \mu \Omega_z}{\Delta(\mu-m_z)} \quad , \\
  \zeta={}& \Gamma_1 \Gamma_2= \frac{\mu+m_z}{\mu-m_z} \quad .
\end{align*}

\subsection{Junctions with in-plane effective magnetization}
In the normal region, where $\Delta=0$, the wave vector component perpendicular to the interface becomes $k^{\lessgtr}_{e,h}= k_{e,h}\! \pm \!m_x/v_F $ with $v_Fk_{e,h}\!=\!\sqrt{\left(\mu\pm E\right)^2\!-\!\left(v_F k_y\pm m_y\right)^2}$. 
The superscript $>$ ($<$) labels right (left) movers along the $x$-direction. 
Electrons and holes have an opposite shift by the field. When they are coupled by the superconductor, this shift becomes very important. 
Therefore, the component of the in-plane magnetization perpendicular to the interface, $m_x=m_{\parallel}\cos\alpha$, discriminates between particles that move with or against the stream of Cooper pairs. 
The general solution of the BdG equations for the normal state region is given in \eref{eq:wvfunction_N} with the same wave functions defined in \eref{eq:wvf_N}, but the new phase factors 
\begin{equation*}
 \e^{i\theta_{e,h}}=\frac{v_Fk_{e,h} + i\left(v_Fk_y\pm m_y\right)}{|\mu\pm E|} \quad .
\end{equation*}

In the heavily-doped regime $\mu\gg E,\Delta$, we have $\theta=\sin^{-1}\left(k_y/\mu\right)$, $k_{1,2}\sim k_{Fx}\pm i\kappa_2$, and $-k_{1,2}\sim-k_{Fx}\mp i\kappa_1$, with 
\[
 \kappa_{1,2}=\frac{\sqrt{|\Delta|^2-(E\pm \mbf{n}\cdot\mbf{m})^2}}{v_F\cos\theta} \quad ,
\]
and $k_{Fx}=k_F\cos\theta$. 
Under this approximation, the wave functions in \eref{eq:wvfunction_S} are
\begin{subequations}
\begin{align}
 \psi_{\pm k_1}={}& \left[ 1, \pm\e^{\pm i\theta}, \mp\Gamma_{\mp}\e^{\pm i\theta}, \Gamma_{\mp} \right]^T \quad , \\
 \psi_{\pm k_2}={}& \left[ \Gamma_{\mp}, \pm\Gamma_{\mp}\e^{\pm i\theta}, \mp\e^{\pm i\theta}, 1 \right]^T \quad .
\end{align}
\label{eq:wvf_Sxy}
\end{subequations}
with
\begin{equation*}
 \Gamma_{\pm}=\frac{\Delta^*}{(E\pm \mbf{n}\cdot\mbf{m})+\sqrt{(E\pm \mbf{n}\cdot\mbf{m})^2-|\Delta|^2}}
 \quad .
\end{equation*} 


\section{Green function techniques for Dirac systems \label{sec:app2}}
For an incoming electron from the normal region, the wave function is
\begin{equation}\label{eq:scattering_1}
 \Psi_1(x)\!\!=\!\! \left\{\!\! \begin{array}{lr}
\begin{array}{c} \e^{ik_e x}\psi_{+k_e} \!+\! a_1 \e^{ik_h x}\psi_{+k_h} \\+ b_1 \e^{-ik_e x}\psi_{-k_e}\end{array} \, , & x<0 
\\ c_1 \e^{ik_1 x}\psi_{+k_1} \!+\! d_1 \e^{-ik_2 x}\psi_{-k_2} \, ,
& x>d
                    \end{array}
\right. \, .
\end{equation}
For the other processes, we find
\begin{equation}\label{eq:scattering_2}
 \Psi_2(x)\!\!=\!\! \left\{\!\! \begin{array}{lr}
\begin{array}{c} \e^{-ik_h x}\psi_{-k_h} \!+\! a_2 \e^{-ik_e x}\psi_{-k_e} \\+ b_2 \e^{ik_h x}\psi_{+k_h} \end{array} \, , & x<0 
\\ c_2 \e^{-ik_2 x}\psi_{-k_2} \!+\! d_2 \e^{ik_1 x}\psi_{+k_1} \, ,
& x>d
                    \end{array}
\right. \, ,
\end{equation}
\begin{equation}\label{eq:scattering_3}
 \Psi_3(x)\!\!=\!\! \left\{\!\! \begin{array}{lr}
c_3 \e^{-ik_e x}\psi_{-k_e} \!+\! d_3 \e^{ik_h x}\psi_{+k_h} \, , & x<0 
\\ \begin{array}{c} \e^{-ik_1 x}\psi_{-k_1} \!+\! a_3 \e^{-ik_2 x}\psi_{-k_2} \\+ b_3 \e^{ik_1 x}\psi_{+k_1} \end{array} \, ,
& x>d
                    \end{array}
\right. \, ,
\end{equation}
and, 
\begin{equation}\label{eq:scattering_4}
 \Psi_4(x)\!\!=\!\! \left\{\!\! \begin{array}{lr}
c_4 \e^{ik_h x}\psi_{+k_h} \!+\! d_3 \e^{-ik_e x}\psi_{-k_e} \, , & x<0 
\\ \begin{array}{c} \e^{ik_2 x}\psi_{+k_2} \!+\! a_4 \e^{ik_1 x}\psi_{+k_1} \\+ b_4 \e^{-ik_2 x}\psi_{-k_2} \end{array} \, ,
& x>d
                    \end{array}
\right. \, .
\end{equation}
At the intermediate region, we use the wave functions
\begin{align}\label{eq:scattering_i}
 \Psi_{j}(0<x<d) ={}&  p_j \e^{ik_e^Ix}\psi_{+k_e^I}^{I} \!+\! q_j\e^{-ik_e^Ix}\psi_{-k_e^I}^{I} \\ & + r_j \e^{-ik_h^Ix}\psi_{-k_h^I}^{I} \!+\! s_j\e^{ik_h^Ix}\psi_{+k_h^I}^{I} \quad , \nonumber 
\end{align}
which correspond to the normal region solutions with the change $\mu\rightarrow\mu_I$ and where $j=1,\dots,4$ labels the processes. 

The corresponding reflection and transmission amplitudes are obtained inserting Eqs. (\ref{eq:wvf_N}), (\ref{eq:wvf_Sz}), and (\ref{eq:wvf_Sxy}) into the boundary conditions
\begin{equation}\label{eq:bbcc}
 \Psi_{j}(0^-)=\Psi_{j}(0^+) \quad , \quad \Psi_{j}(d-0^+)=\Psi_{j}(d+0^+) \quad .
\end{equation}

Next, we insert the resulting wave functions for the scattering processes in the retarded Green function defined in \eref{eq:GF-gen}. The wave functions $\tilde{\Psi}_i(x)$ correspond to the conjugate scattering processes to $\Psi_i(x)$ and are solutions of $\check{H}^*(\mbf{k})$ with the change $\mbf{k}\rightarrow-\mbf{k}$. 
The conjugated wave vectors are obtained by the transformation 
\begin{equation*}
 \tilde{\psi}_{\pm k_x}(k_y)=\psi_{\mp k_x}(-k_y) \quad , 
\end{equation*}
which is a parity transformation. 

We now consider the Green function of the normal region, imposing that $k_{e,h}>0$. 
We substitute the solutions with $x<0$ of Eqs. (\ref{eq:scattering_1}), (\ref{eq:scattering_2}), (\ref{eq:scattering_3}), and (\ref{eq:scattering_4}) into \eref{eq:GF-gen} and apply the continuity condition of \eref{eq:GF-cont} to obtain
\begin{widetext}
 \begin{gather}
  G^r(x,x')= \frac{i}{2v_F} \nonumber \\ \times \left\{\!\! \begin{array}{lr}
                                  \left(\!\!\begin{array}{cc}
\frac{\e^{ik_e|x-x'|}}{\cos\theta_e} \psi_{-k_e}\psi_{-k_e}^{\dagger} + b_1 \frac{\e^{-ik_e(x+x')}}{\cos\theta_e} \psi_{-k_e}\psi_{+k_e}^{\dagger}  & 
a_1 \frac{\e^{-i(k_e x-k_h x')}}{\cos\theta_e} \psi_{-k_e}\psi_{-k_h}^{\dagger} \\
a_2 \frac{\e^{-i(k_e x'-k_h x)}}{\cos\theta_h} \psi_{+k_h}\psi_{+k_e}^{\dagger} &
\frac{\e^{-ik_h|x-x'|}}{\cos\theta_h} \psi_{-k_h}\psi_{-k_h}^{\dagger} + b_2 \frac{\e^{ik_h(x+x')}}{\cos\theta_h} \psi_{+k_h}\psi_{-k_h}^{\dagger}
                                  \end{array}\!\!\right) \,\, ,
& 0>x'>x \\
                                  \left(\!\!\begin{array}{cc}
\frac{\e^{ik_e|x-x'|}}{\cos\theta_e} \psi_{+k_e}\psi_{+k_e}^{\dagger} + b_1 \frac{\e^{-ik_e(x+x')}}{\cos\theta_e} \psi_{-k_e}\psi_{+k_e}^{\dagger} & 
a_2 \frac{\e^{-i(k_e x-k_h x')}}{\cos\theta_h} \psi_{-k_e}\psi_{-k_h}^{\dagger} \\
a_1 \frac{\e^{-i(k_e x'-k_h x)}}{\cos\theta_e} \psi_{+k_h}\psi_{+k_e}^{\dagger} &
\frac{\e^{-ik_h|x-x'|}}{\cos\theta_h} \psi_{-k_h}\psi_{-k_h}^{\dagger} + b_2 \frac{\e^{ik_h(x+x')}}{\cos\theta_h} \psi_{+k_h}\psi_{-k_h}^{\dagger} 
                                  \end{array}\!\!\right) \,\, ,
& 0>x>x' 
                                 \end{array} \right. \quad .
\label{eq:GF-N}
 \end{gather}
Finally, the Green function on the superconducting side of the junction, with $\real\left\{k_{1,2}\right\}>0$,  is 
\begin{gather}
 G^r(x,x')= \nonumber \\ \left\{ \begin{array}{cr}
  \begin{array}{c} 
  A_1 \left[ \e^{ik_1(x'-x)} \psi_{-k_1}\tilde{\psi}_{+k_1}^T  
  + a_3 \e^{i(k_1 x'-k_2 x)}\psi_{-k_2}\tilde{\psi}_{+k_1}^T + b_3 \e^{ik_1(x'+x)} \psi_{+k_1}\tilde{\psi}_{+k_1}^T\right] \\ 
  + A_2 \left[ \e^{-ik_2(x'-x)} \psi_{+k_2}\tilde{\psi}_{-k_2}^T 
  + a_4 \e^{-i(k_2 x'-k_1 x)}\psi_{+k_1}\tilde{\psi}_{-k_2}^T + b_4 \e^{-ik_2(x'+x)}\psi_{-k_2}\tilde{\psi}_{-k_1}^T \right] \end{array} & , x'>x>d \\
  \begin{array}{c} 
  B_1 \left[ \e^{ik_1(x-x')} \psi_{+k_1}\tilde{\psi}_{-k_1}^T 
  + \tilde{a}_3 \e^{i(k_1 x-k_2 x')}\psi_{+k_1}\tilde{\psi}_{-k_2}^T + \tilde{b}_3  \e^{ik_1(x+x')}\psi_{+k_1}\tilde{\psi}_{+k_1}^T \right] \\ 
  + B_2 \left[ \e^{-ik_2(x-x')} \psi_{-k_2}\tilde{\psi}_{+k_2}^T  + \tilde{a}_4 \e^{-i(k_2 x-k_1 x')}\psi_{-k_2}\tilde{\psi}_{+k_1}^T + \tilde{b}_4 \e^{-ik_2(x+x')}\psi_{-k_2}\tilde{\psi}_{-k_2}^T\right] \end{array} & , x>x'>d \\
                       \end{array}
\right. \quad , \label{eq:GF-S}
\end{gather}
\end{widetext}
where
\begin{align*}
 A_1=B_1={}& \frac{i}{v_F}\frac{Z_{1-}}{k_1}\frac{\Gamma_1^{-1}}{\Gamma_1-\Gamma_2} \quad , \\
 A_2=B_2={}& \frac{i}{v_F}\frac{Z_{2-}}{k_2}\frac{\Gamma_2^{-1}}{\Gamma_1-\Gamma_2} \quad , 
\end{align*}
for an out-of-plane magnetization and 
\begin{align*}
 A_1=B_2={}& \frac{i}{v_F\cos\theta}\frac{1}{1-\Gamma_+^2} \quad , \\
 A_2=B_1={}& \frac{i}{v_F\cos\theta}\frac{1}{1-\Gamma_-^2} \quad , 
\end{align*}
for the in-plane magnetization in the heavily-doped, weak-field approximation ($\mu\gg\Delta,E,m_{\parallel}$) and with $|\theta|\leq\pi/2$. For the Green function with $\real\left\{k_{1,2}\right\}<0$, the definition of the coefficients $A_{1,2}$ and $B_{1,2}$ changes sign. Consequently, the sign change in $1/k_{1,2}$ or, equivalently, $1/\cos\theta$ with $\pi/2<\theta<3\pi/2$, is canceled. 

The Green function can be trivially separated into a \textit{bulk} contribution, defined far away from the interface, and an \textit{edge} term which contains the scattering at the interface. This is, $G^r(x,x')\equiv G^r_{\infty}(x,x')+G^r_{A}(x,x')+G^r_{B}(x,x')$, where we have divided the edge term into the part that is given by Andreev reflection processes [$G^r_{A}(x,x')$] and the one that is given by normal reflections [$G^r_{B}(x,x')$].

\begin{figure*}
	\includegraphics[width=\textwidth]{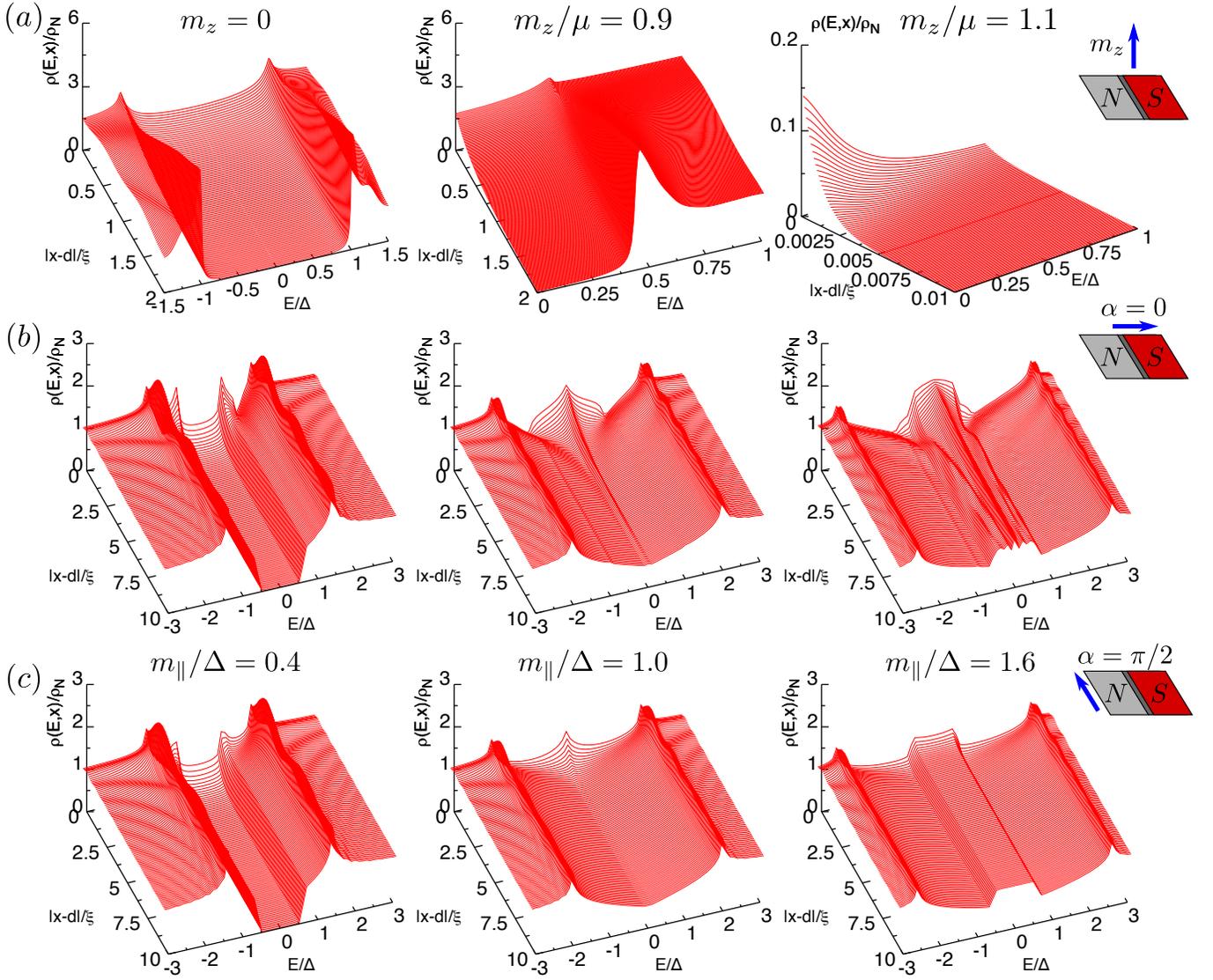}
	\caption{\label{fig:ldos_proximity}
	 LDOS inside the superconducting region normalized to the bulk density of states in the normal region $\rho_N$ as a function of both the excitation energy and the distance from the interface ($x=d$). (a) Out-of-plane magnetization. From left to right, $m_z/\mu=0,0.9,1.1$. For the three panels, $\mu_N=10\mu$ and $\mu_I=20\mu$. (b) In-plane magnetization perpendicular to the NIS interface ($\alpha=0$). From left to right, $m_{\parallel}/\Delta=0.4,1.0,1.6$. (c) In-plane magnetization parallel to the NIS interface ($\alpha=\pi/2$). From left to right, $m_{\parallel}/\Delta=0.4,1.0,1.6$. For (b) and (c), $\mu_N=\mu$ and $\mu_I=10\mu$. For all plots, $\mu=10^{3}\Delta$ and $d=0.1(v_F/\mu)$. }
\end{figure*}

\section{Local density of states \label{sec:app3}}
The electronic LDOS is obtained from the retarded Green function using \eref{eq:ldos-def}. In the normal region is thus given by
\begin{gather}
 \rho_{N}(E)\equiv\rho(E,x\rightarrow-\infty) \nonumber \\ =-\frac{1}{\pi v_F}\int\limits_{-\pi/2}^{\pi/2}\mathrm{d}\theta\cos\theta\imag\left\{\frac{i}{\cos\theta_e}\right\} \quad ,
\end{gather}
for the \textit{bulk} and 
\begin{gather}
 \rho_{B}(E,x)=-\frac{1}{\pi v_F}\int\limits_{\pi/2}^{\pi/2}\mathrm{d}\theta\cos\theta \nonumber \\ \times\imag\left\{\tilde{b}_{1}(E,\theta)\tan\theta_{e}\e^{-i\theta_{e}}\e^{-2ik_{e}x}\right\} \quad ,
\end{gather}
for the \textit{edge}. The LDOS in the normal region is thus $\rho(E,x<0)=\rho_{N}(E)+\rho_{B}(E,x)$. 

Analogously, in the superconducting region we find, in the absence of magnetization, 
\begin{gather}\label{eq:dos_S}
 \rho(E,x>d)=-\frac{1}{\pi v_F}\int\limits_{-\pi/2}^{\pi/2}\mathrm{d}\theta \cos\theta  \\ \imag\left\{ \frac{i\sgn\left(E\right)}{\cos\theta} \left(\frac{1+\Gamma^2}{1-\Gamma^2} +\frac{2\Gamma}{1-\Gamma^2}a_3\e^{-2\kappa x} \right) \right. \nonumber \\ \left. + \frac{\tan\theta}{1-\Gamma^2} \left( \e^{i\theta}\e^{2ik_1 x}b_3 - \Gamma^2\e^{-i\theta}\e^{-2ik_2 x}b_4 \right) \right\} \quad . \nonumber
\end{gather}
We have neglected the rapidly oscillating terms proportional to the normal reflection amplitudes $b_{3,4}(E,\theta)$ in the edge contribution. 
The \textit{bulk} contribution of \eref{eq:dos_S} reduces to the BCS density of states $\rho(E,\mbf{m}=0)=\real\left\{E/\sqrt{E^2-\Delta^2}\right\}$. We plot in the left panel of \fref{fig:ldos_proximity}(a) the LDOS for zero field as a function of the energy and the distance from the interface inside the superconducting region. The LDOS is finite for $|E|\le\Delta$ at the interface ($x=d$) and decays to zero for distances inside the superconducting region comparable to the superconducting coherence length. 

When we consider an in-plane magnetization, the previous result is changed to
\begin{gather}
 \rho(E,x>d)=-\frac{1}{\pi v_F}\int\limits_{-\pi/2}^{\pi/2}\mathrm{d}\theta \cos\theta  \\ \imag\left( \frac{i\sgn\left(E\right)}{\cos\theta} \left[\frac{1-\Gamma_{+}^2\Gamma_{-}^2}{(1-\Gamma_{+}^2)(1-\Gamma_{-}^2)} \right. \right. \nonumber \\ \left. \left. +\left(\frac{\Gamma_{+}}{1-\Gamma_{+}^2}a_3+\frac{\Gamma_{-}}{1-\Gamma_{-}^2}a_4\right)\e^{-2\kappa x} \right] + \tan\theta \right. \nonumber \\ \left. \times \left\{ \begin{array}{lr} \frac{\e^{i\theta}b_3}{1-\Gamma_{+}^2} \e^{2ik_1 x} - \frac{\Gamma_{+}\Gamma_{-}}{1-\Gamma_{-}^2} \e^{-i\theta}\e^{-2ik_2 x}b_4 & , E\ge0 \\ \frac{\e^{-i\theta}b_3}{1-\Gamma_{-}^2} \e^{2ik_1 x} - \frac{\Gamma_{+}\Gamma_{-}}{1-\Gamma_{+}^2} \e^{i\theta}\e^{-2ik_2 x}b_4 & , E<0 \end{array} \right. \right) \quad . \nonumber
\label{eq:dos_mxy_S}
\end{gather}
We plot in \fref{fig:ldos_proximity}(b,c) the LDOS as a function of the energy and the distance from the IS interface at $x=d$ for several values of $m_{\parallel}$ and for $\alpha=0$ and $\alpha=\pi/2$, respectively. The effect of the in-plane magnetization is to split the superconducting gap into two, $\Delta\rightarrow\Delta_{\pm}$. When $m_{\parallel}<\Delta$, the LDOS at the interface clearly shows four resonances at $|E|=\Delta_{\pm}$ which become, inside the superconducting region, fully gapped for $|E|\le\Delta_{-}$ with sharp resonances at $|E|=\Delta_{+}$ [see left panels of \fref{fig:ldos_proximity}(b,c)]. For $m_{\parallel}=\Delta$, $\Delta_{-}=0$ and the LDOS inside the superconducting region adopts a V-shaped profile. At the interface, a peak at $E=0$ appears which has a long-range decay if $\alpha=0$ ($|x-d|\sim10\xi$). For a magnetization oriented parallel to the interface ($\alpha=\pi/2$), the peak decays as $|x-d|\sim\xi$. This peak disappears when $m_{\parallel}>\Delta$ if $\alpha>\pi/4$, but it is still present and displays a long-range decay otherwise. 

Finally, for an out-of-plane magnetization, the LDOS is given by
\begin{gather}
 \rho(E,x>d)=-\frac{1}{\pi v_F}\int\mathrm{d}k_y \imag\left\{ \sgn\left(E\right) \right. \\ \left. \times \left[ A_1\Gamma_1^2(1+E_{1+}E_{1-})+A_2\Gamma_2^2(1+E_{2+}E_{2-}) \right. \right. \nonumber \\ \left. \left. +   \Gamma_1\Gamma_2(A_1a_2+A_2a_4)\e^{-2\kappa_z x} \right. \right. \nonumber \\ \left. \left. \times \left\{ \begin{array}{lr} 1+E_{1+}E_{2-} & , E\ge0 \\ 1+E_{1-}E_{2+} & , E<0 \end{array}\right\} \right. \right. \nonumber \\ \left. \left. + A_1\Gamma_1^2(1-E_{1+}^2)\e^{2ik_1 x} b_3 \right.\right. \nonumber \\ \left.\left. - A_2\Gamma_2^2(1-E_{2-}^2)\e^{-2ik_2 x} b_4 \right] \right\} \quad . \nonumber
\label{eq:dos_mz_S}
\end{gather}
We show the LDOS results for a perpendicular magnetization close to the closing of the gap in the central and right panels of \fref{fig:ldos_proximity}(a). Before the closing of the gap (central panel with $m_z=0.9$), the LDOS shows a resonance at the energies corresponding to the effective gap $\Delta_z$ and is greatly enhanced at zero energy close to the interface. This enhancement of the LDOS, however, decays fast inside the superconducting region and disappears at a distance comparable to the superconducting coherence length. After closing the gap (right panel with $m_z=1.1$), superconductivity is strongly suppressed and only a zero-energy peak on the LDOS survives at the interface. This peak decays inside the superconducting region within $10^{-2}\xi$. 

\section{Anomalous Green function \label{sec:app4}}
We now analyze the electron-hole component in Nambu space of \eref{eq:GF-S}. 
For simplicity, we only consider NS junctions with no intermediate region. Moreover, we only show the case with $\real\left\{k_x\right\}>0$. The symmetry classification remains the same for more complicated NIS junctions. 

In the absence of magnetization, the Andreev reflection probabilities adopt the simple form $a_3=a_4=\tilde{a}_3=\tilde{a}_4=-\Gamma$ and the normal reflections are $b_3=b_4=\tilde{b}_3=\tilde{b}_4=0$. The components of the anomalous Green function are thus given by 
\begin{align*}
 F_0(E,\theta)\!={}& \frac{i\sgn\left(E\right)}{2v_F\cos\theta} \left[ \frac{2\Gamma}{1-\Gamma^2} + \frac{1+\Gamma^2}{1-\Gamma^2} a_3\e^{-2\kappa x}\right] \, , \\
 F_3(E,\theta)\!={}& 0 \, , \\
 F_{\up\up}(E,\theta)\!={}&F_{\dw\dw}(E,\theta) \nonumber \\ ={}&\frac{\sgn\left(E\right)}{2v_F\cos\theta} \left[ \frac{2\Gamma}{1-\Gamma^2} + \frac{1+\Gamma^2}{1-\Gamma^2} a_3\e^{-2\kappa x}\right]\sin\theta \nonumber \\&+ \frac{i\tilde{\sigma}}{2v_F}a_3\e^{-2\kappa x} \quad , 
\end{align*}
with $\tilde{\sigma}=+1,-1$ for $F_{\up\up}$ and $F_{\dw\dw}$, respectively. We have defined $\Gamma=\Delta/(E+\Omega)$, $\kappa=-i\Omega/(v_F\cos\theta)$, and $\Omega=\sqrt{E^2-\Delta^2}$. 
The singlet term $F_0(E,\theta)$ is even in frequency and spatial dependence; therefore, it is classified as ESE. The first term of $F_{\sigma\sigma}(E,\theta)$, with $\sigma=\up,\dw$, is equal to $F_0(E,\theta)$ multiplied by $\sin\theta$, which makes it even in energy but odd in spatial dependence, hence classified as ETO. The second term of $F_{\sigma\sigma}(E,\theta)$ is odd in energy and even in spatial dependence and is classified as OTE. 

In the main text, we use the triplet components $F_{1,2}=F_{\up\up}\mp F_{\dw\dw}$. Due to the sign change for the \textit{edge} part proportional to the Andreev reflection amplitude $a_3$, $F_2$ is equal to the \textit{bulk} ETO term multiplied by $\sin\theta$ while $F_1$ is given by the \text{edge} OTE part. 
After averaging over the angle of incidence, only the ESE and OTE terms are non-zero and we obtain the behavior shown in \fref{fig:fdos_z}(a) of the main text. 

We now consider an in-plane magnetization. For NS junctions with no intermediate region, we find $a_3=\tilde{a}_4=-\Gamma_+$, $a_4=\tilde{a}_3=-\Gamma_-$, and $b_3=b_4=\tilde{b}_3=\tilde{b}_4=0$. The singlet component is 
\begin{gather}
 F_0(E,\theta)\!= \frac{i\sgn\left(E\right)}{2v_F\cos\theta} \!\left(\! \frac{\Gamma_+}{1-\Gamma_+^2}\!+\!\frac{\Gamma_-}{1-\Gamma_-^2} \!\right)\! +\! \frac{i\e^{-(\kappa_1-\kappa_2) x}}{2v_F\cos\theta} \nonumber \\ \times \left\{\begin{array}{cr} +\left(\frac{\Gamma_+^2}{1-\Gamma_+^2} a_3+\frac{1}{1-\Gamma_-^2} a_4\right) \quad , & E\ge0 \\ -\left(\frac{1}{1-\Gamma_+^2} a_3+\frac{\Gamma_-^2}{1-\Gamma_-^2} a_4\right) \quad , & E<0 \end{array}\right.  \, ,
\end{gather}
which reduces to the previous result for $m_{\parallel}=0$, where $\Gamma_+=\Gamma_-=\Gamma$, and is still classified as ESE. 
For the triplet components, we find that $F_3(E,\theta)=0$ and 
\begin{gather}
 F_{\sigma\sigma}(E,\theta) \!\!=\!\! \left[-\frac{\sgn\left(E\right)}{2\cos\theta}\!\left(\! \frac{\Gamma_+}{1-\Gamma_+^2}\!+\!\frac{\Gamma_-}{1-\Gamma_-^2}\! \right) \!+\! \frac{\e^{-(\kappa_1-\kappa_2) x}}{2v_F\cos\theta} \right. \nonumber \\ \left. \times \left\{\begin{array}{cr} +\left(\frac{\Gamma_+^2}{1-\Gamma_+^2} a_3+\frac{1}{1-\Gamma_-^2} a_4\right) \quad , & E\ge0 \\ -\left(\frac{1}{1-\Gamma_+^2} a_3+\frac{\Gamma_-^2}{1-\Gamma_-^2} a_4\right) \quad , & E<0 \end{array}\right\} \right]\sin\theta \nonumber \\ \times \frac{i\tilde{\sigma}}{2v_F}\left(\! \frac{\Gamma_+}{1-\Gamma_+^2}\!-\!\frac{\Gamma_-}{1-\Gamma_-^2}\! \right) + \frac{i\tilde{\sigma}\e^{-(\kappa_1-\kappa_2) x}}{2v_F} \nonumber \\ \times \left\{\begin{array}{cr} +\left(\frac{\Gamma_+^2}{1-\Gamma_+^2} a_3-\frac{1}{1-\Gamma_-^2} a_4\right) \quad , & E\ge0 \\ -\left(\frac{1}{1-\Gamma_+^2} a_3-\frac{\Gamma_-^2}{1-\Gamma_-^2} a_4\right) \quad , & E<0 \end{array}\right.  \, ,
\end{gather}
As before, the term proportional to $\sin\theta$ is even in energy, thus belonging to ETO classification, while the term proportional to $\tilde{\sigma}$ is odd in energy and is classified as OTE. 

Finally, we consider the out-of-plane magnetization. In this case, the reflection amplitudes adopt a rather complicated form. For simplicity, in the following analysis we only consider the terms proportional to Andreev reflection amplitudes, which fulfill $a_i=\tilde{a}_i$. The terms coming from normal reflections only add some rapid spatial oscillations. Under these approximations, the spin-singlet component of the anomalous Green function is
\begin{gather}
  F_0(E,k_y)= \frac{i\sgn\left(E\right)}{4v_F\left(\Gamma_1 \!-\! \Gamma_2\right)}  \!\left(\frac{Z_{1+} + \zeta Z_{1-}}{k_1}\!+\!\frac{Z_{2+} + \zeta Z_{2-}}{k_2}\right) \nonumber \\ + \frac{i\e^{-2\kappa_z x}}{4v_F\left(\Gamma_1-\Gamma_2\right)}\left( \frac{Z_{1+}\Gamma_2}{\Gamma_1k_1}a_3 + \zeta\frac{Z_{2-}\Gamma_1}{\Gamma_2k_2}a_4 \right) \nonumber \\ \times \sgn\left(E\right) \left[ 1+\frac{k_1k_2}{Z_{1+}Z_{2-}} - ik_y\frac{k_1-k_2}{Z_{1+}Z_{2-}}\right] \quad .
\end{gather}
The \textit{bulk} part of the singlet component is even in energy and belongs to ESE classification. For the \textit{edge} part, proportional to $\exp(-2\kappa_z x)$, with $\kappa_z=\mu\Omega_z/v_F\sqrt{\mu^2-m_z^2-k_y^2}\approx-i(k_1-k_2)/2$, we find another ESE term together with an OSO term. The latter, proportional to $k_y$, is only present when there is electron-hole asymmetry and vanishes if $\mu\gg E,\Delta$. 

When the out-of-plane magnetization is finite, we find a new triplet component, which was zero in the previous analysis, namely,
\begin{gather}
  F_3(E,k_y)= -\frac{i\sgn\left(E\right)}{2v_F\left(\Gamma_1-\Gamma_2\right)} \frac{Em_z}{\mu-m_z}\left(\frac{1}{k_1}\!+\!\frac{1}{k_2}\right) \nonumber \\ + \frac{i\e^{-2\kappa_z x}}{4v_F\left(\Gamma_1-\Gamma_2\right)}\left( \frac{Z_{1+}\Gamma_2}{\Gamma_1k_1}a_3 + \zeta\frac{Z_{2-}\Gamma_1}{\Gamma_2k_2}a_4 \right) \nonumber \\ \times \left[ \sgn\left(E\right)\left(1-\frac{k_1k_2}{Z_{1+}Z_{2-}}\right) - ik_y\frac{k_1-k_2}{Z_{1+}Z_{2-}} \right] \quad .
\end{gather}
The \textit{bulk} part, which classifies as OTE, is zero for $E=0$. The \textit{edge} part has components from both ETO and OTE. For the other triplet components, we find
\begin{align}
  F_{\up\up}(E,k_y)={}& -k_y\sgn\left(E\right)\zeta\frac{k_1+k_2}{2v_F\left(\Gamma_1-\Gamma_2\right)} \\ &+ \frac{i\zeta\e^{-2\kappa_z x}}{4v_F\left(\Gamma_1-\Gamma_2\right)} \left[ \left( \frac{\Gamma_2}{\Gamma_1}a_3 - \frac{\Gamma_1}{\Gamma_2}a_4 \right) \right. \nonumber \\ &\left. + i k_y\sgn\left(E\right)\left( \frac{\Gamma_2}{\Gamma_1k_1}a_3 + \zeta\frac{\Gamma_1}{\Gamma_2k_2}a_4 \right) \right] \quad , \nonumber 
\end{align}
and
\begin{gather}
  F_{\dw\dw}(E,k_y)= -k_y\sgn\left(E\right)\frac{k_1+k_2}{2v_F\left(\Gamma_1-\Gamma_2\right)} \\ - \frac{i\e^{-2\kappa_z x}}{4v_F\left(\Gamma_1-\Gamma_2\right)} \left[ \left( \frac{Z_{1+}}{\zeta Z_{2-}} \frac{\Gamma_2k_2}{\Gamma_1k_1}a_3 - \frac{\zeta Z_{2-}}{Z_{1+}} \frac{\Gamma_1k_1}{\Gamma_2k_2}a_4 \right) \right. \nonumber \\ \left. + i k_y\sgn\left(E\right)\left( \frac{Z_{1+}}{\zeta Z_{2-}} \frac{\Gamma_2}{\Gamma_1k_1}a_3 - \frac{\zeta Z_{2-}}{Z_{1+}} \frac{\Gamma_1}{\Gamma_2k_2}a_4 \right) \right] \quad . \nonumber
\end{gather}
As described in the main text, both \textit{bulk} terms have ETO symmetry and are canceled after averaging over incident angles. 

%

\begin{thebibliography}{71}%
	\makeatletter
	\providecommand \@ifxundefined [1]{%
		\@ifx{#1\undefined}
	}%
	\providecommand \@ifnum [1]{%
		\ifnum #1\expandafter \@firstoftwo
		\else \expandafter \@secondoftwo
		\fi
	}%
	\providecommand \@ifx [1]{%
		\ifx #1\expandafter \@firstoftwo
		\else \expandafter \@secondoftwo
		\fi
	}%
	\providecommand \natexlab [1]{#1}%
	\providecommand \enquote  [1]{``#1''}%
	\providecommand \bibnamefont  [1]{#1}%
	\providecommand \bibfnamefont [1]{#1}%
	\providecommand \citenamefont [1]{#1}%
	\providecommand \href@noop [0]{\@secondoftwo}%
	\providecommand \href [0]{\begingroup \@sanitize@url \@href}%
	\providecommand \@href[1]{\@@startlink{#1}\@@href}%
	\providecommand \@@href[1]{\endgroup#1\@@endlink}%
	\providecommand \@sanitize@url [0]{\catcode `\\12\catcode `\$12\catcode
		`\&12\catcode `\#12\catcode `\^12\catcode `\_12\catcode `\%12\relax}%
	\providecommand \@@startlink[1]{}%
	\providecommand \@@endlink[0]{}%
	\providecommand \url  [0]{\begingroup\@sanitize@url \@url }%
	\providecommand \@url [1]{\endgroup\@href {#1}{\urlprefix }}%
	\providecommand \urlprefix  [0]{URL }%
	\providecommand \Eprint [0]{\href }%
	\providecommand \doibase [0]{http://dx.doi.org/}%
	\providecommand \selectlanguage [0]{\@gobble}%
	\providecommand \bibinfo  [0]{\@secondoftwo}%
	\providecommand \bibfield  [0]{\@secondoftwo}%
	\providecommand \translation [1]{[#1]}%
	\providecommand \BibitemOpen [0]{}%
	\providecommand \bibitemStop [0]{}%
	\providecommand \bibitemNoStop [0]{.\EOS\space}%
	\providecommand \EOS [0]{\spacefactor3000\relax}%
	\providecommand \BibitemShut  [1]{\csname bibitem#1\endcsname}%
	\let\auto@bib@innerbib\@empty
	\bibitem [{\citenamefont {Qi}\ and\ \citenamefont {Zhang}(2011)}]{Qi_RMP}%
	\BibitemOpen
	\bibfield  {author} {\bibinfo {author} {\bibfnamefont {X.-L.}\ \bibnamefont
			{Qi}}\ and\ \bibinfo {author} {\bibfnamefont {S.-C.}\ \bibnamefont {Zhang}},\
	}\href {\doibase 10.1103/RevModPhys.83.1057} {\bibfield  {journal} {\bibinfo
		{journal} {Rev. Mod. Phys.}\ }\textbf {\bibinfo {volume} {83}},\ \bibinfo
	{pages} {1057} (\bibinfo {year} {2011})}\BibitemShut {NoStop}%
\bibitem [{\citenamefont {Alicea}(2012)}]{Alicea_Majorana}%
\BibitemOpen
\bibfield  {author} {\bibinfo {author} {\bibfnamefont {J.}~\bibnamefont
		{Alicea}},\ }\href {http://stacks.iop.org/0034-4885/75/i=7/a=076501}
{\bibfield  {journal} {\bibinfo  {journal} {Reports on Progress in Physics}\
	}\textbf {\bibinfo {volume} {75}},\ \bibinfo {pages} {076501} (\bibinfo
	{year} {2012})}\BibitemShut {NoStop}%
\bibitem [{\citenamefont {Beenakker}(2013)}]{Beenakker_Majorana}%
\BibitemOpen
\bibfield  {author} {\bibinfo {author} {\bibfnamefont {C.}~\bibnamefont
		{Beenakker}},\ }\href {\doibase 10.1146/annurev-conmatphys-030212-184337}
{\bibfield  {journal} {\bibinfo  {journal} {Annual Review of Condensed Matter
			Physics}\ }\textbf {\bibinfo {volume} {4}},\ \bibinfo {pages} {113} (\bibinfo
	{year} {2013})}\BibitemShut {NoStop}%
\bibitem [{\citenamefont {Fu}\ and\ \citenamefont {Kane}(2008)}]{Fu_2008}%
\BibitemOpen
\bibfield  {author} {\bibinfo {author} {\bibfnamefont {L.}~\bibnamefont
		{Fu}}\ and\ \bibinfo {author} {\bibfnamefont {C.~L.}\ \bibnamefont {Kane}},\
}\href {\doibase 10.1103/PhysRevLett.100.096407} {\bibfield  {journal}
{\bibinfo  {journal} {Phys. Rev. Lett.}\ }\textbf {\bibinfo {volume} {100}},\
\bibinfo {pages} {096407} (\bibinfo {year} {2008})}\BibitemShut {NoStop}%
\bibitem [{\citenamefont {Sau}\ \emph {et~al.}(2010)\citenamefont {Sau},
	\citenamefont {Lutchyn}, \citenamefont {Tewari},\ and\ \citenamefont
	{Das~Sarma}}]{Sau_2010}%
\BibitemOpen
\bibfield  {author} {\bibinfo {author} {\bibfnamefont {J.~D.}\ \bibnamefont
		{Sau}}, \bibinfo {author} {\bibfnamefont {R.~M.}\ \bibnamefont {Lutchyn}},
	\bibinfo {author} {\bibfnamefont {S.}~\bibnamefont {Tewari}}, \ and\ \bibinfo
	{author} {\bibfnamefont {S.}~\bibnamefont {Das~Sarma}},\ }\href {\doibase
	10.1103/PhysRevLett.104.040502} {\bibfield  {journal} {\bibinfo  {journal}
		{Phys. Rev. Lett.}\ }\textbf {\bibinfo {volume} {104}},\ \bibinfo {pages}
	{040502} (\bibinfo {year} {2010})}\BibitemShut {NoStop}%
\bibitem [{\citenamefont {Lutchyn}\ \emph {et~al.}(2010)\citenamefont
	{Lutchyn}, \citenamefont {Sau},\ and\ \citenamefont
	{Das~Sarma}}]{Lutchyn_2010}%
\BibitemOpen
\bibfield  {author} {\bibinfo {author} {\bibfnamefont {R.~M.}\ \bibnamefont
		{Lutchyn}}, \bibinfo {author} {\bibfnamefont {J.~D.}\ \bibnamefont {Sau}}, \
	and\ \bibinfo {author} {\bibfnamefont {S.}~\bibnamefont {Das~Sarma}},\ }\href
{\doibase 10.1103/PhysRevLett.105.077001} {\bibfield  {journal} {\bibinfo
		{journal} {Phys. Rev. Lett.}\ }\textbf {\bibinfo {volume} {105}},\ \bibinfo
	{pages} {077001} (\bibinfo {year} {2010})}\BibitemShut {NoStop}%
\bibitem [{\citenamefont {Alicea}(2010)}]{Alicea_2010}%
\BibitemOpen
\bibfield  {author} {\bibinfo {author} {\bibfnamefont {J.}~\bibnamefont
		{Alicea}},\ }\href {\doibase 10.1103/PhysRevB.81.125318} {\bibfield
	{journal} {\bibinfo  {journal} {Phys. Rev. B}\ }\textbf {\bibinfo {volume}
		{81}},\ \bibinfo {pages} {125318} (\bibinfo {year} {2010})}\BibitemShut
{NoStop}%
\bibitem [{\citenamefont {Oreg}\ \emph {et~al.}(2010)\citenamefont {Oreg},
	\citenamefont {Refael},\ and\ \citenamefont {von Oppen}}]{Oreg_2010}%
\BibitemOpen
\bibfield  {author} {\bibinfo {author} {\bibfnamefont {Y.}~\bibnamefont
		{Oreg}}, \bibinfo {author} {\bibfnamefont {G.}~\bibnamefont {Refael}}, \ and\
	\bibinfo {author} {\bibfnamefont {F.}~\bibnamefont {von Oppen}},\ }\href
{\doibase 10.1103/PhysRevLett.105.177002} {\bibfield  {journal} {\bibinfo
		{journal} {Phys. Rev. Lett.}\ }\textbf {\bibinfo {volume} {105}},\ \bibinfo
	{pages} {177002} (\bibinfo {year} {2010})}\BibitemShut {NoStop}%
\bibitem [{\citenamefont {Potter}\ and\ \citenamefont
	{Lee}(2010)}]{Potter_2010}%
\BibitemOpen
\bibfield  {author} {\bibinfo {author} {\bibfnamefont {A.~C.}\ \bibnamefont
		{Potter}}\ and\ \bibinfo {author} {\bibfnamefont {P.~A.}\ \bibnamefont
		{Lee}},\ }\href {\doibase 10.1103/PhysRevLett.105.227003} {\bibfield
	{journal} {\bibinfo  {journal} {Phys. Rev. Lett.}\ }\textbf {\bibinfo
		{volume} {105}},\ \bibinfo {pages} {227003} (\bibinfo {year}
	{2010})}\BibitemShut {NoStop}%
\bibitem [{\citenamefont {Potter}\ and\ \citenamefont
	{Lee}(2011)}]{Potter_2011}%
\BibitemOpen
\bibfield  {author} {\bibinfo {author} {\bibfnamefont {A.~C.}\ \bibnamefont
		{Potter}}\ and\ \bibinfo {author} {\bibfnamefont {P.~A.}\ \bibnamefont
		{Lee}},\ }\href {\doibase 10.1103/PhysRevB.83.184520} {\bibfield  {journal}
	{\bibinfo  {journal} {Phys. Rev. B}\ }\textbf {\bibinfo {volume} {83}},\
	\bibinfo {pages} {184520} (\bibinfo {year} {2011})}\BibitemShut {NoStop}%
\bibitem [{\citenamefont {Prada}\ \emph {et~al.}(2012)\citenamefont {Prada},
	\citenamefont {San-Jose},\ and\ \citenamefont {Aguado}}]{Prada_2012}%
\BibitemOpen
\bibfield  {author} {\bibinfo {author} {\bibfnamefont {E.}~\bibnamefont
		{Prada}}, \bibinfo {author} {\bibfnamefont {P.}~\bibnamefont {San-Jose}}, \
	and\ \bibinfo {author} {\bibfnamefont {R.}~\bibnamefont {Aguado}},\ }\href
{\doibase 10.1103/PhysRevB.86.180503} {\bibfield  {journal} {\bibinfo
		{journal} {Phys. Rev. B}\ }\textbf {\bibinfo {volume} {86}},\ \bibinfo
	{pages} {180503} (\bibinfo {year} {2012})}\BibitemShut {NoStop}%
\bibitem [{\citenamefont {Houzet}\ \emph {et~al.}(2013)\citenamefont {Houzet},
	\citenamefont {Meyer}, \citenamefont {Badiane},\ and\ \citenamefont
	{Glazman}}]{Houzet_2013}%
\BibitemOpen
\bibfield  {author} {\bibinfo {author} {\bibfnamefont {M.}~\bibnamefont
		{Houzet}}, \bibinfo {author} {\bibfnamefont {J.~S.}\ \bibnamefont {Meyer}},
	\bibinfo {author} {\bibfnamefont {D.~M.}\ \bibnamefont {Badiane}}, \ and\
	\bibinfo {author} {\bibfnamefont {L.~I.}\ \bibnamefont {Glazman}},\ }\href
{\doibase 10.1103/PhysRevLett.111.046401} {\bibfield  {journal} {\bibinfo
		{journal} {Phys. Rev. Lett.}\ }\textbf {\bibinfo {volume} {111}},\ \bibinfo
	{pages} {046401} (\bibinfo {year} {2013})}\BibitemShut {NoStop}%
\bibitem [{\citenamefont {Tiwari}\ \emph {et~al.}(2013)\citenamefont {Tiwari},
	\citenamefont {Z\"ulicke},\ and\ \citenamefont {Bruder}}]{Tiwari_2013}%
\BibitemOpen
\bibfield  {author} {\bibinfo {author} {\bibfnamefont {R.~P.}\ \bibnamefont
		{Tiwari}}, \bibinfo {author} {\bibfnamefont {U.}~\bibnamefont {Z\"ulicke}}, \
	and\ \bibinfo {author} {\bibfnamefont {C.}~\bibnamefont {Bruder}},\ }\href
{\doibase 10.1103/PhysRevLett.110.186805} {\bibfield  {journal} {\bibinfo
		{journal} {Phys. Rev. Lett.}\ }\textbf {\bibinfo {volume} {110}},\ \bibinfo
	{pages} {186805} (\bibinfo {year} {2013})}\BibitemShut {NoStop}%
\bibitem [{\citenamefont {Tiwari}\ \emph {et~al.}(2014)\citenamefont {Tiwari},
	\citenamefont {Z\"ulicke},\ and\ \citenamefont {Bruder}}]{Tiwari_2014b}%
\BibitemOpen
\bibfield  {author} {\bibinfo {author} {\bibfnamefont {R.~P.}\ \bibnamefont
		{Tiwari}}, \bibinfo {author} {\bibfnamefont {U.}~\bibnamefont {Z\"ulicke}}, \
	and\ \bibinfo {author} {\bibfnamefont {C.}~\bibnamefont {Bruder}},\ }\href
{http://stacks.iop.org/1367-2630/16/i=2/a=025004} {\bibfield  {journal}
	{\bibinfo  {journal} {New Journal of Physics}\ }\textbf {\bibinfo {volume}
		{16}},\ \bibinfo {pages} {025004} (\bibinfo {year} {2014})}\BibitemShut
{NoStop}%
\bibitem [{\citenamefont {Rex}\ and\ \citenamefont
	{Sudb\o{}}(2014)}]{Rex_2014}%
\BibitemOpen
\bibfield  {author} {\bibinfo {author} {\bibfnamefont {S.}~\bibnamefont
		{Rex}}\ and\ \bibinfo {author} {\bibfnamefont {A.}~\bibnamefont {Sudb\o{}}},\
}\href {\doibase 10.1103/PhysRevB.90.115429} {\bibfield  {journal} {\bibinfo
	{journal} {Phys. Rev. B}\ }\textbf {\bibinfo {volume} {90}},\ \bibinfo
{pages} {115429} (\bibinfo {year} {2014})}\BibitemShut {NoStop}%
\bibitem [{\citenamefont {Read}\ and\ \citenamefont {Green}(2000)}]{Read_2000}%
\BibitemOpen
\bibfield  {author} {\bibinfo {author} {\bibfnamefont {N.}~\bibnamefont
		{Read}}\ and\ \bibinfo {author} {\bibfnamefont {D.}~\bibnamefont {Green}},\
}\href {\doibase 10.1103/PhysRevB.61.10267} {\bibfield  {journal} {\bibinfo
	{journal} {Phys. Rev. B}\ }\textbf {\bibinfo {volume} {61}},\ \bibinfo
{pages} {10267} (\bibinfo {year} {2000})}\BibitemShut {NoStop}%
\bibitem [{\citenamefont {Gor'kov}\ and\ \citenamefont
	{Rashba}(2001)}]{Gorkov_2001}%
\BibitemOpen
\bibfield  {author} {\bibinfo {author} {\bibfnamefont {L.~P.}\ \bibnamefont
		{Gor'kov}}\ and\ \bibinfo {author} {\bibfnamefont {E.~I.}\ \bibnamefont
		{Rashba}},\ }\href {\doibase 10.1103/PhysRevLett.87.037004} {\bibfield
	{journal} {\bibinfo  {journal} {Phys. Rev. Lett.}\ }\textbf {\bibinfo
		{volume} {87}},\ \bibinfo {pages} {037004} (\bibinfo {year}
	{2001})}\BibitemShut {NoStop}%
\bibitem [{\citenamefont {Schnyder}\ \emph {et~al.}(2008)\citenamefont
	{Schnyder}, \citenamefont {Ryu}, \citenamefont {Furusaki},\ and\
	\citenamefont {Ludwig}}]{Schnyder_2008}%
\BibitemOpen
\bibfield  {author} {\bibinfo {author} {\bibfnamefont {A.~P.}\ \bibnamefont
		{Schnyder}}, \bibinfo {author} {\bibfnamefont {S.}~\bibnamefont {Ryu}},
	\bibinfo {author} {\bibfnamefont {A.}~\bibnamefont {Furusaki}}, \ and\
	\bibinfo {author} {\bibfnamefont {A.~W.~W.}\ \bibnamefont {Ludwig}},\ }\href
{\doibase 10.1103/PhysRevB.78.195125} {\bibfield  {journal} {\bibinfo
		{journal} {Phys. Rev. B}\ }\textbf {\bibinfo {volume} {78}},\ \bibinfo
	{pages} {195125} (\bibinfo {year} {2008})}\BibitemShut {NoStop}%
\bibitem [{\citenamefont {Yokoyama}(2012)}]{Yokoyama_2012}%
\BibitemOpen
\bibfield  {author} {\bibinfo {author} {\bibfnamefont {T.}~\bibnamefont
		{Yokoyama}},\ }\href {\doibase 10.1103/PhysRevB.86.075410} {\bibfield
	{journal} {\bibinfo  {journal} {Phys. Rev. B}\ }\textbf {\bibinfo {volume}
		{86}},\ \bibinfo {pages} {075410} (\bibinfo {year} {2012})}\BibitemShut
{NoStop}%
\bibitem [{\citenamefont {Tkachov}\ and\ \citenamefont
	{Hankiewicz}(2013{\natexlab{a}})}]{Tkachov_2012}%
\BibitemOpen
\bibfield  {author} {\bibinfo {author} {\bibfnamefont {G.}~\bibnamefont
		{Tkachov}}\ and\ \bibinfo {author} {\bibfnamefont {E.~M.}\ \bibnamefont
		{Hankiewicz}},\ }\href {\doibase 10.1002/pssb.201248385} {\bibfield
	{journal} {\bibinfo  {journal} {Phys. Status Solidi B}\ }\textbf {\bibinfo
		{volume} {250}},\ \bibinfo {pages} {215} (\bibinfo {year}
	{2013}{\natexlab{a}})}\BibitemShut {NoStop}%
\bibitem [{\citenamefont {Tkachov}(2013)}]{Tkachov_2013}%
\BibitemOpen
\bibfield  {author} {\bibinfo {author} {\bibfnamefont {G.}~\bibnamefont
		{Tkachov}},\ }\href {\doibase 10.1103/PhysRevB.87.245422} {\bibfield
	{journal} {\bibinfo  {journal} {Phys. Rev. B}\ }\textbf {\bibinfo {volume}
		{87}},\ \bibinfo {pages} {245422} (\bibinfo {year} {2013})}\BibitemShut
{NoStop}%
\bibitem [{\citenamefont {Tkachov}\ and\ \citenamefont
	{Hankiewicz}(2013{\natexlab{b}})}]{Tkachov_2013b}%
\BibitemOpen
\bibfield  {author} {\bibinfo {author} {\bibfnamefont {G.}~\bibnamefont
		{Tkachov}}\ and\ \bibinfo {author} {\bibfnamefont {E.~M.}\ \bibnamefont
		{Hankiewicz}},\ }\href {\doibase 10.1103/PhysRevB.88.075401} {\bibfield
	{journal} {\bibinfo  {journal} {Phys. Rev. B}\ }\textbf {\bibinfo {volume}
		{88}},\ \bibinfo {pages} {075401} (\bibinfo {year}
	{2013}{\natexlab{b}})}\BibitemShut {NoStop}%
\bibitem [{\citenamefont {Burset}\ \emph {et~al.}(2014)\citenamefont {Burset},
	\citenamefont {Keidel}, \citenamefont {Tanaka}, \citenamefont {Nagaosa},\
	and\ \citenamefont {Trauzettel}}]{Burset_2014}%
\BibitemOpen
\bibfield  {author} {\bibinfo {author} {\bibfnamefont {P.}~\bibnamefont
		{Burset}}, \bibinfo {author} {\bibfnamefont {F.}~\bibnamefont {Keidel}},
	\bibinfo {author} {\bibfnamefont {Y.}~\bibnamefont {Tanaka}}, \bibinfo
	{author} {\bibfnamefont {N.}~\bibnamefont {Nagaosa}}, \ and\ \bibinfo
	{author} {\bibfnamefont {B.}~\bibnamefont {Trauzettel}},\ }\href {\doibase
	10.1103/PhysRevB.90.085438} {\bibfield  {journal} {\bibinfo  {journal} {Phys.
			Rev. B}\ }\textbf {\bibinfo {volume} {90}},\ \bibinfo {pages} {085438}
	(\bibinfo {year} {2014})}\BibitemShut {NoStop}%
\bibitem [{\citenamefont {Tanaka}\ and\ \citenamefont
	{Golubov}(2007)}]{Tanaka_2007}%
\BibitemOpen
\bibfield  {author} {\bibinfo {author} {\bibfnamefont {Y.}~\bibnamefont
		{Tanaka}}\ and\ \bibinfo {author} {\bibfnamefont {A.~A.}\ \bibnamefont
		{Golubov}},\ }\href {\doibase 10.1103/PhysRevLett.98.037003} {\bibfield
	{journal} {\bibinfo  {journal} {Phys. Rev. Lett.}\ }\textbf {\bibinfo
		{volume} {98}},\ \bibinfo {pages} {037003} (\bibinfo {year}
	{2007})}\BibitemShut {NoStop}%
\bibitem [{\citenamefont {Eschrig}\ \emph {et~al.}(2007)\citenamefont
	{Eschrig}, \citenamefont {L\"ofwander}, \citenamefont {Champel},
	\citenamefont {Cuevas}, \citenamefont {Kopu},\ and\ \citenamefont
	{Sch\"on}}]{Eschrig_2007}%
\BibitemOpen
\bibfield  {author} {\bibinfo {author} {\bibfnamefont {M.}~\bibnamefont
		{Eschrig}}, \bibinfo {author} {\bibfnamefont {T.}~\bibnamefont
		{L\"ofwander}}, \bibinfo {author} {\bibfnamefont {T.}~\bibnamefont
		{Champel}}, \bibinfo {author} {\bibfnamefont {J.}~\bibnamefont {Cuevas}},
	\bibinfo {author} {\bibfnamefont {J.}~\bibnamefont {Kopu}}, \ and\ \bibinfo
	{author} {\bibfnamefont {G.}~\bibnamefont {Sch\"on}},\ }\href {\doibase
	10.1007/s10909-007-9329-6} {\bibfield  {journal} {\bibinfo  {journal}
		{Journal of Low Temperature Physics}\ }\textbf {\bibinfo {volume} {147}},\
	\bibinfo {pages} {457} (\bibinfo {year} {2007})}\BibitemShut {NoStop}%
\bibitem [{\citenamefont {Tanaka}\ \emph
	{et~al.}(2007{\natexlab{a}})\citenamefont {Tanaka}, \citenamefont {Golubov},
	\citenamefont {Kashiwaya},\ and\ \citenamefont {Ueda}}]{Tanaka_2007b}%
\BibitemOpen
\bibfield  {author} {\bibinfo {author} {\bibfnamefont {Y.}~\bibnamefont
		{Tanaka}}, \bibinfo {author} {\bibfnamefont {A.~A.}\ \bibnamefont {Golubov}},
	\bibinfo {author} {\bibfnamefont {S.}~\bibnamefont {Kashiwaya}}, \ and\
	\bibinfo {author} {\bibfnamefont {M.}~\bibnamefont {Ueda}},\ }\href {\doibase
	10.1103/PhysRevLett.99.037005} {\bibfield  {journal} {\bibinfo  {journal}
		{Phys. Rev. Lett.}\ }\textbf {\bibinfo {volume} {99}},\ \bibinfo {pages}
	{037005} (\bibinfo {year} {2007}{\natexlab{a}})}\BibitemShut {NoStop}%
\bibitem [{\citenamefont {Tanaka}\ \emph
	{et~al.}(2007{\natexlab{b}})\citenamefont {Tanaka}, \citenamefont {Tanuma},\
	and\ \citenamefont {Golubov}}]{Tanaka_2007c}%
\BibitemOpen
\bibfield  {author} {\bibinfo {author} {\bibfnamefont {Y.}~\bibnamefont
		{Tanaka}}, \bibinfo {author} {\bibfnamefont {Y.}~\bibnamefont {Tanuma}}, \
	and\ \bibinfo {author} {\bibfnamefont {A.~A.}\ \bibnamefont {Golubov}},\
}\href {\doibase 10.1103/PhysRevB.76.054522} {\bibfield  {journal} {\bibinfo
	{journal} {Phys. Rev. B}\ }\textbf {\bibinfo {volume} {76}},\ \bibinfo
{pages} {054522} (\bibinfo {year} {2007}{\natexlab{b}})}\BibitemShut
{NoStop}%
\bibitem [{\citenamefont {Tanaka}\ \emph {et~al.}(2012)\citenamefont {Tanaka},
	\citenamefont {Sato},\ and\ \citenamefont {Nagaosa}}]{Tanaka_JPSJ}%
\BibitemOpen
\bibfield  {author} {\bibinfo {author} {\bibfnamefont {Y.}~\bibnamefont
		{Tanaka}}, \bibinfo {author} {\bibfnamefont {M.}~\bibnamefont {Sato}}, \ and\
	\bibinfo {author} {\bibfnamefont {N.}~\bibnamefont {Nagaosa}},\ }\href
{\doibase 10.1143/JPSJ.81.011013} {\bibfield  {journal} {\bibinfo  {journal}
		{Journal of the Physical Society of Japan}\ }\textbf {\bibinfo {volume}
		{81}},\ \bibinfo {pages} {011013} (\bibinfo {year} {2012})}\BibitemShut
{NoStop}%
\bibitem [{\citenamefont {Black-Schaffer}\ and\ \citenamefont
	{Balatsky}(2012)}]{Black-Schaffer_2012}%
\BibitemOpen
\bibfield  {author} {\bibinfo {author} {\bibfnamefont {A.~M.}\ \bibnamefont
		{Black-Schaffer}}\ and\ \bibinfo {author} {\bibfnamefont {A.~V.}\
		\bibnamefont {Balatsky}},\ }\href {\doibase 10.1103/PhysRevB.86.144506}
{\bibfield  {journal} {\bibinfo  {journal} {Phys. Rev. B}\ }\textbf {\bibinfo
		{volume} {86}},\ \bibinfo {pages} {144506} (\bibinfo {year}
	{2012})}\BibitemShut {NoStop}%
\bibitem [{\citenamefont {Black-Schaffer}\ and\ \citenamefont
	{Balatsky}(2013{\natexlab{a}})}]{Black-Schaffer_2013}%
\BibitemOpen
\bibfield  {author} {\bibinfo {author} {\bibfnamefont {A.~M.}\ \bibnamefont
		{Black-Schaffer}}\ and\ \bibinfo {author} {\bibfnamefont {A.~V.}\
		\bibnamefont {Balatsky}},\ }\href {\doibase 10.1103/PhysRevB.87.220506}
{\bibfield  {journal} {\bibinfo  {journal} {Phys. Rev. B}\ }\textbf {\bibinfo
		{volume} {87}},\ \bibinfo {pages} {220506} (\bibinfo {year}
	{2013}{\natexlab{a}})}\BibitemShut {NoStop}%
\bibitem [{\citenamefont {Black-Schaffer}\ and\ \citenamefont
	{Balatsky}(2013{\natexlab{b}})}]{Black-Schaffer_2013b}%
\BibitemOpen
\bibfield  {author} {\bibinfo {author} {\bibfnamefont {A.~M.}\ \bibnamefont
		{Black-Schaffer}}\ and\ \bibinfo {author} {\bibfnamefont {A.~V.}\
		\bibnamefont {Balatsky}},\ }\href {\doibase 10.1103/PhysRevB.88.104514}
{\bibfield  {journal} {\bibinfo  {journal} {Phys. Rev. B}\ }\textbf {\bibinfo
		{volume} {88}},\ \bibinfo {pages} {104514} (\bibinfo {year}
	{2013}{\natexlab{b}})}\BibitemShut {NoStop}%
\bibitem [{\citenamefont {Berezinskii}(1974{\natexlab{a}})}]{Berezinskii_1974}%
\BibitemOpen
\bibfield  {author} {\bibinfo {author} {\bibfnamefont {V.~L.}\ \bibnamefont
		{Berezinskii}},\ }\href@noop {} {\bibfield  {journal} {\bibinfo  {journal}
		{JETP Letters}\ }\textbf {\bibinfo {volume} {20}},\ \bibinfo {pages} {287}
	(\bibinfo {year} {1974}{\natexlab{a}})}\BibitemShut {NoStop}%
\bibitem [{\citenamefont
	{Berezinskii}(1974{\natexlab{b}})}]{Berezinskii_1974R}%
\BibitemOpen
\bibfield  {author} {\bibinfo {author} {\bibfnamefont {V.~L.}\ \bibnamefont
		{Berezinskii}},\ }\href@noop {} {\bibfield  {journal} {\bibinfo  {journal}
		{Pis’ma Zh. Eksp. Teor. Fiz.}\ }\textbf {\bibinfo {volume} {20}},\ \bibinfo
	{pages} {628} (\bibinfo {year} {1974}{\natexlab{b}})}\BibitemShut {NoStop}%
\bibitem [{\citenamefont {Balatsky}\ and\ \citenamefont
	{Abrahams}(1992)}]{Balatsky_1992}%
\BibitemOpen
\bibfield  {author} {\bibinfo {author} {\bibfnamefont {A.}~\bibnamefont
		{Balatsky}}\ and\ \bibinfo {author} {\bibfnamefont {E.}~\bibnamefont
		{Abrahams}},\ }\href {\doibase 10.1103/PhysRevB.45.13125} {\bibfield
	{journal} {\bibinfo  {journal} {Phys. Rev. B}\ }\textbf {\bibinfo {volume}
		{45}},\ \bibinfo {pages} {13125} (\bibinfo {year} {1992})}\BibitemShut
{NoStop}%
\bibitem [{\citenamefont {Bergeret}\ \emph {et~al.}(2001)\citenamefont
	{Bergeret}, \citenamefont {Volkov},\ and\ \citenamefont
	{Efetov}}]{Bergeret_2001}%
\BibitemOpen
\bibfield  {author} {\bibinfo {author} {\bibfnamefont {F.~S.}\ \bibnamefont
		{Bergeret}}, \bibinfo {author} {\bibfnamefont {A.~F.}\ \bibnamefont
		{Volkov}}, \ and\ \bibinfo {author} {\bibfnamefont {K.~B.}\ \bibnamefont
		{Efetov}},\ }\href {\doibase 10.1103/PhysRevLett.86.4096} {\bibfield
	{journal} {\bibinfo  {journal} {Phys. Rev. Lett.}\ }\textbf {\bibinfo
		{volume} {86}},\ \bibinfo {pages} {4096} (\bibinfo {year}
	{2001})}\BibitemShut {NoStop}%
\bibitem [{\citenamefont {Khaire}\ \emph {et~al.}(2010)\citenamefont {Khaire},
	\citenamefont {Khasawneh}, \citenamefont {Pratt},\ and\ \citenamefont
	{Birge}}]{Birge_2010}%
\BibitemOpen
\bibfield  {author} {\bibinfo {author} {\bibfnamefont {T.~S.}\ \bibnamefont
		{Khaire}}, \bibinfo {author} {\bibfnamefont {M.~A.}\ \bibnamefont
		{Khasawneh}}, \bibinfo {author} {\bibfnamefont {W.~P.}\ \bibnamefont
		{Pratt}}, \ and\ \bibinfo {author} {\bibfnamefont {N.~O.}\ \bibnamefont
		{Birge}},\ }\href {\doibase 10.1103/PhysRevLett.104.137002} {\bibfield
	{journal} {\bibinfo  {journal} {Phys. Rev. Lett.}\ }\textbf {\bibinfo
		{volume} {104}},\ \bibinfo {pages} {137002} (\bibinfo {year}
	{2010})}\BibitemShut {NoStop}%
\bibitem [{\citenamefont {Bergeret}\ \emph {et~al.}(2005)\citenamefont
	{Bergeret}, \citenamefont {Volkov},\ and\ \citenamefont
	{Efetov}}]{Bergeret_RMP}%
\BibitemOpen
\bibfield  {author} {\bibinfo {author} {\bibfnamefont {F.~S.}\ \bibnamefont
		{Bergeret}}, \bibinfo {author} {\bibfnamefont {A.~F.}\ \bibnamefont
		{Volkov}}, \ and\ \bibinfo {author} {\bibfnamefont {K.~B.}\ \bibnamefont
		{Efetov}},\ }\href {\doibase 10.1103/RevModPhys.77.1321} {\bibfield
	{journal} {\bibinfo  {journal} {Rev. Mod. Phys.}\ }\textbf {\bibinfo {volume}
		{77}},\ \bibinfo {pages} {1321} (\bibinfo {year} {2005})}\BibitemShut
{NoStop}%
\bibitem [{\citenamefont {Eschrig}(2010)}]{Eschrig_2010}%
\BibitemOpen
\bibfield  {author} {\bibinfo {author} {\bibfnamefont {M.}~\bibnamefont
		{Eschrig}},\ }\href {http://dx.doi.org/10.1063/1.3541944} {\bibfield
	{journal} {\bibinfo  {journal} {Phys. Today}\ }\textbf {\bibinfo {volume}
		{64}},\ \bibinfo {pages} {43} (\bibinfo {year} {2010})}\BibitemShut {NoStop}%
\bibitem [{\citenamefont {Cr\'epin}\ \emph {et~al.}(2015)\citenamefont
	{Cr\'epin}, \citenamefont {Burset},\ and\ \citenamefont
	{Trauzettel}}]{Crepin_2015}%
\BibitemOpen
\bibfield  {author} {\bibinfo {author} {\bibfnamefont {F.}~\bibnamefont
		{Cr\'epin}}, \bibinfo {author} {\bibfnamefont {P.}~\bibnamefont {Burset}}, \
	and\ \bibinfo {author} {\bibfnamefont {B.}~\bibnamefont {Trauzettel}},\
}\href {\doibase 10.1103/PhysRevB.92.100507} {\bibfield  {journal} {\bibinfo
	{journal} {Phys. Rev. B}\ }\textbf {\bibinfo {volume} {92}},\ \bibinfo
{pages} {100507} (\bibinfo {year} {2015})}\BibitemShut {NoStop}%
\bibitem [{\citenamefont {Asano}\ and\ \citenamefont
	{Tanaka}(2013)}]{Asano_2013}%
\BibitemOpen
\bibfield  {author} {\bibinfo {author} {\bibfnamefont {Y.}~\bibnamefont
		{Asano}}\ and\ \bibinfo {author} {\bibfnamefont {Y.}~\bibnamefont {Tanaka}},\
}\href {\doibase 10.1103/PhysRevB.87.104513} {\bibfield  {journal} {\bibinfo
	{journal} {Phys. Rev. B}\ }\textbf {\bibinfo {volume} {87}},\ \bibinfo
{pages} {104513} (\bibinfo {year} {2013})}\BibitemShut {NoStop}%
\bibitem [{\citenamefont {Stanev}\ and\ \citenamefont
	{Galitski}(2014)}]{Galitski_2014}%
\BibitemOpen
\bibfield  {author} {\bibinfo {author} {\bibfnamefont {V.}~\bibnamefont
		{Stanev}}\ and\ \bibinfo {author} {\bibfnamefont {V.}~\bibnamefont
		{Galitski}},\ }\href {\doibase 10.1103/PhysRevB.89.174521} {\bibfield
	{journal} {\bibinfo  {journal} {Phys. Rev. B}\ }\textbf {\bibinfo {volume}
		{89}},\ \bibinfo {pages} {174521} (\bibinfo {year} {2014})}\BibitemShut
{NoStop}%
\bibitem [{\citenamefont {Liu}\ \emph {et~al.}(2015)\citenamefont {Liu},
	\citenamefont {Sau},\ and\ \citenamefont {Das~Sarma}}]{DasSarma_2015}%
\BibitemOpen
\bibfield  {author} {\bibinfo {author} {\bibfnamefont {X.}~\bibnamefont
		{Liu}}, \bibinfo {author} {\bibfnamefont {J.~D.}\ \bibnamefont {Sau}}, \ and\
	\bibinfo {author} {\bibfnamefont {S.}~\bibnamefont {Das~Sarma}},\ }\href
{\doibase 10.1103/PhysRevB.92.014513} {\bibfield  {journal} {\bibinfo
		{journal} {Phys. Rev. B}\ }\textbf {\bibinfo {volume} {92}},\ \bibinfo
	{pages} {014513} (\bibinfo {year} {2015})}\BibitemShut {NoStop}%
\bibitem [{\citenamefont {Ebisu}\ \emph {et~al.}(2015)\citenamefont {Ebisu},
	\citenamefont {Yada}, \citenamefont {Kasai},\ and\ \citenamefont
	{Tanaka}}]{Ebisu_2015}%
\BibitemOpen
\bibfield  {author} {\bibinfo {author} {\bibfnamefont {H.}~\bibnamefont
		{Ebisu}}, \bibinfo {author} {\bibfnamefont {K.}~\bibnamefont {Yada}},
	\bibinfo {author} {\bibfnamefont {H.}~\bibnamefont {Kasai}}, \ and\ \bibinfo
	{author} {\bibfnamefont {Y.}~\bibnamefont {Tanaka}},\ }\href {\doibase
	10.1103/PhysRevB.91.054518} {\bibfield  {journal} {\bibinfo  {journal} {Phys.
			Rev. B}\ }\textbf {\bibinfo {volume} {91}},\ \bibinfo {pages} {054518}
	(\bibinfo {year} {2015})}\BibitemShut {NoStop}%
\bibitem [{\citenamefont {Snelder}\ \emph
	{et~al.}(2015{\natexlab{a}})\citenamefont {Snelder}, \citenamefont {Golubov},
	\citenamefont {Asano},\ and\ \citenamefont {Brinkman}}]{Snelder_2015}%
\BibitemOpen
\bibfield  {author} {\bibinfo {author} {\bibfnamefont {M.}~\bibnamefont
		{Snelder}}, \bibinfo {author} {\bibfnamefont {A.~A.}\ \bibnamefont
		{Golubov}}, \bibinfo {author} {\bibfnamefont {Y.}~\bibnamefont {Asano}}, \
	and\ \bibinfo {author} {\bibfnamefont {A.}~\bibnamefont {Brinkman}},\ }\href
{http://stacks.iop.org/0953-8984/27/i=31/a=315701} {\bibfield  {journal}
	{\bibinfo  {journal} {Journal of Physics: Condensed Matter}\ }\textbf
	{\bibinfo {volume} {27}},\ \bibinfo {pages} {315701} (\bibinfo {year}
	{2015}{\natexlab{a}})}\BibitemShut {NoStop}%
\bibitem [{\citenamefont {Tanaka}\ \emph
	{et~al.}(2009{\natexlab{a}})\citenamefont {Tanaka}, \citenamefont
	{Yokoyama},\ and\ \citenamefont {Nagaosa}}]{Tanaka_2009}%
\BibitemOpen
\bibfield  {author} {\bibinfo {author} {\bibfnamefont {Y.}~\bibnamefont
		{Tanaka}}, \bibinfo {author} {\bibfnamefont {T.}~\bibnamefont {Yokoyama}}, \
	and\ \bibinfo {author} {\bibfnamefont {N.}~\bibnamefont {Nagaosa}},\ }\href
{\doibase 10.1103/PhysRevLett.103.107002} {\bibfield  {journal} {\bibinfo
		{journal} {Phys. Rev. Lett.}\ }\textbf {\bibinfo {volume} {103}},\ \bibinfo
	{pages} {107002} (\bibinfo {year} {2009}{\natexlab{a}})}\BibitemShut
{NoStop}%
\bibitem [{\citenamefont {Linder}\ \emph
	{et~al.}(2010{\natexlab{a}})\citenamefont {Linder}, \citenamefont {Tanaka},
	\citenamefont {Yokoyama}, \citenamefont {Sudb\o{}},\ and\ \citenamefont
	{Nagaosa}}]{Linder_2010b}%
\BibitemOpen
\bibfield  {author} {\bibinfo {author} {\bibfnamefont {J.}~\bibnamefont
		{Linder}}, \bibinfo {author} {\bibfnamefont {Y.}~\bibnamefont {Tanaka}},
	\bibinfo {author} {\bibfnamefont {T.}~\bibnamefont {Yokoyama}}, \bibinfo
	{author} {\bibfnamefont {A.}~\bibnamefont {Sudb\o{}}}, \ and\ \bibinfo
	{author} {\bibfnamefont {N.}~\bibnamefont {Nagaosa}},\ }\href {\doibase
	10.1103/PhysRevB.81.184525} {\bibfield  {journal} {\bibinfo  {journal} {Phys.
			Rev. B}\ }\textbf {\bibinfo {volume} {81}},\ \bibinfo {pages} {184525}
	(\bibinfo {year} {2010}{\natexlab{a}})}\BibitemShut {NoStop}%
\bibitem [{\citenamefont {Soori}\ \emph {et~al.}(2013)\citenamefont {Soori},
	\citenamefont {Deb}, \citenamefont {Sengupta},\ and\ \citenamefont
	{Sen}}]{Sengupta_2013}%
\BibitemOpen
\bibfield  {author} {\bibinfo {author} {\bibfnamefont {A.}~\bibnamefont
		{Soori}}, \bibinfo {author} {\bibfnamefont {O.}~\bibnamefont {Deb}}, \bibinfo
	{author} {\bibfnamefont {K.}~\bibnamefont {Sengupta}}, \ and\ \bibinfo
	{author} {\bibfnamefont {D.}~\bibnamefont {Sen}},\ }\href {\doibase
	10.1103/PhysRevB.87.245435} {\bibfield  {journal} {\bibinfo  {journal} {Phys.
			Rev. B}\ }\textbf {\bibinfo {volume} {87}},\ \bibinfo {pages} {245435}
	(\bibinfo {year} {2013})}\BibitemShut {NoStop}%
\bibitem [{\citenamefont {Linder}\ \emph
	{et~al.}(2010{\natexlab{b}})\citenamefont {Linder}, \citenamefont {Tanaka},
	\citenamefont {Yokoyama}, \citenamefont {Sudb\o{}},\ and\ \citenamefont
	{Nagaosa}}]{Linder_2010}%
\BibitemOpen
\bibfield  {author} {\bibinfo {author} {\bibfnamefont {J.}~\bibnamefont
		{Linder}}, \bibinfo {author} {\bibfnamefont {Y.}~\bibnamefont {Tanaka}},
	\bibinfo {author} {\bibfnamefont {T.}~\bibnamefont {Yokoyama}}, \bibinfo
	{author} {\bibfnamefont {A.}~\bibnamefont {Sudb\o{}}}, \ and\ \bibinfo
	{author} {\bibfnamefont {N.}~\bibnamefont {Nagaosa}},\ }\href {\doibase
	10.1103/PhysRevLett.104.067001} {\bibfield  {journal} {\bibinfo  {journal}
		{Phys. Rev. Lett.}\ }\textbf {\bibinfo {volume} {104}},\ \bibinfo {pages}
	{067001} (\bibinfo {year} {2010}{\natexlab{b}})}\BibitemShut {NoStop}%
\bibitem [{\citenamefont {Olund}\ and\ \citenamefont
	{Zhao}(2012)}]{Olund_2012}%
\BibitemOpen
\bibfield  {author} {\bibinfo {author} {\bibfnamefont {C.~T.}\ \bibnamefont
		{Olund}}\ and\ \bibinfo {author} {\bibfnamefont {E.}~\bibnamefont {Zhao}},\
}\href {\doibase 10.1103/PhysRevB.86.214515} {\bibfield  {journal} {\bibinfo
	{journal} {Phys. Rev. B}\ }\textbf {\bibinfo {volume} {86}},\ \bibinfo
{pages} {214515} (\bibinfo {year} {2012})}\BibitemShut {NoStop}%
\bibitem [{\citenamefont {Snelder}\ \emph {et~al.}(2013)\citenamefont
	{Snelder}, \citenamefont {Veldhorst}, \citenamefont {Golubov},\ and\
	\citenamefont {Brinkman}}]{Snelder_2013}%
\BibitemOpen
\bibfield  {author} {\bibinfo {author} {\bibfnamefont {M.}~\bibnamefont
		{Snelder}}, \bibinfo {author} {\bibfnamefont {M.}~\bibnamefont {Veldhorst}},
	\bibinfo {author} {\bibfnamefont {A.~A.}\ \bibnamefont {Golubov}}, \ and\
	\bibinfo {author} {\bibfnamefont {A.}~\bibnamefont {Brinkman}},\ }\href
{\doibase 10.1103/PhysRevB.87.104507} {\bibfield  {journal} {\bibinfo
		{journal} {Phys. Rev. B}\ }\textbf {\bibinfo {volume} {87}},\ \bibinfo
	{pages} {104507} (\bibinfo {year} {2013})}\BibitemShut {NoStop}%
\bibitem [{\citenamefont {Nussbaum}\ \emph {et~al.}(2014)\citenamefont
	{Nussbaum}, \citenamefont {Schmidt}, \citenamefont {Bruder},\ and\
	\citenamefont {Tiwari}}]{Tiwari_2014}%
\BibitemOpen
\bibfield  {author} {\bibinfo {author} {\bibfnamefont {J.}~\bibnamefont
		{Nussbaum}}, \bibinfo {author} {\bibfnamefont {T.~L.}\ \bibnamefont
		{Schmidt}}, \bibinfo {author} {\bibfnamefont {C.}~\bibnamefont {Bruder}}, \
	and\ \bibinfo {author} {\bibfnamefont {R.~P.}\ \bibnamefont {Tiwari}},\
}\href {\doibase 10.1103/PhysRevB.90.045413} {\bibfield  {journal} {\bibinfo
	{journal} {Phys. Rev. B}\ }\textbf {\bibinfo {volume} {90}},\ \bibinfo
{pages} {045413} (\bibinfo {year} {2014})}\BibitemShut {NoStop}%
\bibitem [{\citenamefont {Lu}\ \emph {et~al.}(2015{\natexlab{a}})\citenamefont
	{Lu}, \citenamefont {Burset}, \citenamefont {Yada},\ and\ \citenamefont
	{Tanaka}}]{Lu_2015}%
\BibitemOpen
\bibfield  {author} {\bibinfo {author} {\bibfnamefont {B.}~\bibnamefont
		{Lu}}, \bibinfo {author} {\bibfnamefont {P.}~\bibnamefont {Burset}}, \bibinfo
	{author} {\bibfnamefont {K.}~\bibnamefont {Yada}}, \ and\ \bibinfo {author}
	{\bibfnamefont {Y.}~\bibnamefont {Tanaka}},\ }\href
{http://stacks.iop.org/0953-2048/28/i=10/a=105001} {\bibfield  {journal}
	{\bibinfo  {journal} {Superconductor Science and Technology}\ }\textbf
	{\bibinfo {volume} {28}},\ \bibinfo {pages} {105001} (\bibinfo {year}
	{2015}{\natexlab{a}})}\BibitemShut {NoStop}%
\bibitem [{\citenamefont {Lu}\ \emph {et~al.}(2015{\natexlab{b}})\citenamefont
	{Lu}, \citenamefont {Yada}, \citenamefont {Golubov},\ and\ \citenamefont
	{Tanaka}}]{Lu_2015b}%
\BibitemOpen
\bibfield  {author} {\bibinfo {author} {\bibfnamefont {B.}~\bibnamefont
		{Lu}}, \bibinfo {author} {\bibfnamefont {K.}~\bibnamefont {Yada}}, \bibinfo
	{author} {\bibfnamefont {A.~A.}\ \bibnamefont {Golubov}}, \ and\ \bibinfo
	{author} {\bibfnamefont {Y.}~\bibnamefont {Tanaka}},\ }\href {\doibase
	10.1103/PhysRevB.92.100503} {\bibfield  {journal} {\bibinfo  {journal} {Phys.
			Rev. B}\ }\textbf {\bibinfo {volume} {92}},\ \bibinfo {pages} {100503}
	(\bibinfo {year} {2015}{\natexlab{b}})}\BibitemShut {NoStop}%
\bibitem [{\citenamefont {Tkachov}\ \emph {et~al.}(2015)\citenamefont
	{Tkachov}, \citenamefont {Burset}, \citenamefont {Trauzettel},\ and\
	\citenamefont {Hankiewicz}}]{Tkachov_2015}%
\BibitemOpen
\bibfield  {author} {\bibinfo {author} {\bibfnamefont {G.}~\bibnamefont
		{Tkachov}}, \bibinfo {author} {\bibfnamefont {P.}~\bibnamefont {Burset}},
	\bibinfo {author} {\bibfnamefont {B.}~\bibnamefont {Trauzettel}}, \ and\
	\bibinfo {author} {\bibfnamefont {E.~M.}\ \bibnamefont {Hankiewicz}},\ }\href
{\doibase 10.1103/PhysRevB.92.045408} {\bibfield  {journal} {\bibinfo
		{journal} {Phys. Rev. B}\ }\textbf {\bibinfo {volume} {92}},\ \bibinfo
	{pages} {045408} (\bibinfo {year} {2015})}\BibitemShut {NoStop}%
\bibitem [{\citenamefont {Wei}\ \emph {et~al.}(2013)\citenamefont {Wei},
	\citenamefont {Katmis}, \citenamefont {Assaf}, \citenamefont {Steinberg},
	\citenamefont {Jarillo-Herrero}, \citenamefont {Heiman},\ and\ \citenamefont
	{Moodera}}]{Moodera_2013}%
\BibitemOpen
\bibfield  {author} {\bibinfo {author} {\bibfnamefont {P.}~\bibnamefont
		{Wei}}, \bibinfo {author} {\bibfnamefont {F.}~\bibnamefont {Katmis}},
	\bibinfo {author} {\bibfnamefont {B.~A.}\ \bibnamefont {Assaf}}, \bibinfo
	{author} {\bibfnamefont {H.}~\bibnamefont {Steinberg}}, \bibinfo {author}
	{\bibfnamefont {P.}~\bibnamefont {Jarillo-Herrero}}, \bibinfo {author}
	{\bibfnamefont {D.}~\bibnamefont {Heiman}}, \ and\ \bibinfo {author}
	{\bibfnamefont {J.~S.}\ \bibnamefont {Moodera}},\ }\href {\doibase
	10.1103/PhysRevLett.110.186807} {\bibfield  {journal} {\bibinfo  {journal}
		{Phys. Rev. Lett.}\ }\textbf {\bibinfo {volume} {110}},\ \bibinfo {pages}
	{186807} (\bibinfo {year} {2013})}\BibitemShut {NoStop}%
\bibitem [{\citenamefont {Fogelstr\"om}\ \emph {et~al.}(1997)\citenamefont
	{Fogelstr\"om}, \citenamefont {Rainer},\ and\ \citenamefont
	{Sauls}}]{Sauls_1997}%
\BibitemOpen
\bibfield  {author} {\bibinfo {author} {\bibfnamefont {M.}~\bibnamefont
		{Fogelstr\"om}}, \bibinfo {author} {\bibfnamefont {D.}~\bibnamefont
		{Rainer}}, \ and\ \bibinfo {author} {\bibfnamefont {J.~A.}\ \bibnamefont
		{Sauls}},\ }\href {\doibase 10.1103/PhysRevLett.79.281} {\bibfield  {journal}
	{\bibinfo  {journal} {Phys. Rev. Lett.}\ }\textbf {\bibinfo {volume} {79}},\
	\bibinfo {pages} {281} (\bibinfo {year} {1997})}\BibitemShut {NoStop}%
\bibitem [{\citenamefont {Tanaka}\ \emph {et~al.}(2002)\citenamefont {Tanaka},
	\citenamefont {Tanuma}, \citenamefont {Kuroki},\ and\ \citenamefont
	{Kashiwaya}}]{Tanaka_2002}%
\BibitemOpen
\bibfield  {author} {\bibinfo {author} {\bibfnamefont {Y.}~\bibnamefont
		{Tanaka}}, \bibinfo {author} {\bibfnamefont {Y.}~\bibnamefont {Tanuma}},
	\bibinfo {author} {\bibfnamefont {K.}~\bibnamefont {Kuroki}}, \ and\ \bibinfo
	{author} {\bibfnamefont {S.}~\bibnamefont {Kashiwaya}},\ }\href {\doibase
	10.1143/JPSJ.71.2102} {\bibfield  {journal} {\bibinfo  {journal} {Journal of
			the Physical Society of Japan}\ }\textbf {\bibinfo {volume} {71}},\ \bibinfo
	{pages} {2102} (\bibinfo {year} {2002})}\BibitemShut {NoStop}%
\bibitem [{\citenamefont {Tanaka}\ \emph
	{et~al.}(2009{\natexlab{b}})\citenamefont {Tanaka}, \citenamefont {Yokoyama},
	\citenamefont {Balatsky},\ and\ \citenamefont {Nagaosa}}]{Tanaka_2009b}%
\BibitemOpen
\bibfield  {author} {\bibinfo {author} {\bibfnamefont {Y.}~\bibnamefont
		{Tanaka}}, \bibinfo {author} {\bibfnamefont {T.}~\bibnamefont {Yokoyama}},
	\bibinfo {author} {\bibfnamefont {A.~V.}\ \bibnamefont {Balatsky}}, \ and\
	\bibinfo {author} {\bibfnamefont {N.}~\bibnamefont {Nagaosa}},\ }\href
{\doibase 10.1103/PhysRevB.79.060505} {\bibfield  {journal} {\bibinfo
		{journal} {Phys. Rev. B}\ }\textbf {\bibinfo {volume} {79}},\ \bibinfo
	{pages} {060505} (\bibinfo {year} {2009}{\natexlab{b}})}\BibitemShut
{NoStop}%
\bibitem [{\citenamefont {Burset}\ \emph {et~al.}(2009)\citenamefont {Burset},
	\citenamefont {Herrera},\ and\ \citenamefont {Levy~Yeyati}}]{Burset_2009}%
\BibitemOpen
\bibfield  {author} {\bibinfo {author} {\bibfnamefont {P.}~\bibnamefont
		{Burset}}, \bibinfo {author} {\bibfnamefont {W.}~\bibnamefont {Herrera}}, \
	and\ \bibinfo {author} {\bibfnamefont {A.}~\bibnamefont {Levy~Yeyati}},\
}\href {\doibase 10.1103/PhysRevB.80.041402} {\bibfield  {journal} {\bibinfo
	{journal} {Phys. Rev. B}\ }\textbf {\bibinfo {volume} {80}},\ \bibinfo
{pages} {041402} (\bibinfo {year} {2009})}\BibitemShut {NoStop}%
\bibitem [{\citenamefont {McMillan}(1968)}]{McMillan_1968}%
\BibitemOpen
\bibfield  {author} {\bibinfo {author} {\bibfnamefont {W.~L.}\ \bibnamefont
		{McMillan}},\ }\href {\doibase 10.1103/PhysRev.175.559} {\bibfield  {journal}
	{\bibinfo  {journal} {Phys. Rev.}\ }\textbf {\bibinfo {volume} {175}},\
	\bibinfo {pages} {559} (\bibinfo {year} {1968})}\BibitemShut {NoStop}%
\bibitem [{\citenamefont {Furusaki}\ and\ \citenamefont
	{Tsukada}(1991)}]{Furusaki_1991}%
\BibitemOpen
\bibfield  {author} {\bibinfo {author} {\bibfnamefont {A.}~\bibnamefont
		{Furusaki}}\ and\ \bibinfo {author} {\bibfnamefont {M.}~\bibnamefont
		{Tsukada}},\ }\href {\doibase http://dx.doi.org/10.1016/0038-1098(91)90201-6}
{\bibfield  {journal} {\bibinfo  {journal} {Solid State Communications}\
	}\textbf {\bibinfo {volume} {78}},\ \bibinfo {pages} {299 } (\bibinfo {year}
	{1991})}\BibitemShut {NoStop}%
\bibitem [{\citenamefont {Tanaka}\ and\ \citenamefont
	{Kashiwaya}(1996)}]{Tanaka_1996}%
\BibitemOpen
\bibfield  {author} {\bibinfo {author} {\bibfnamefont {Y.}~\bibnamefont
		{Tanaka}}\ and\ \bibinfo {author} {\bibfnamefont {S.}~\bibnamefont
		{Kashiwaya}},\ }\href {\doibase 10.1103/PhysRevB.53.9371} {\bibfield
	{journal} {\bibinfo  {journal} {Phys. Rev. B}\ }\textbf {\bibinfo {volume}
		{53}},\ \bibinfo {pages} {9371} (\bibinfo {year} {1996})}\BibitemShut
{NoStop}%
\bibitem [{\citenamefont {Kashiwaya}\ and\ \citenamefont
	{Tanaka}(2000)}]{Kashiwaya_2000}%
\BibitemOpen
\bibfield  {author} {\bibinfo {author} {\bibfnamefont {S.}~\bibnamefont
		{Kashiwaya}}\ and\ \bibinfo {author} {\bibfnamefont {Y.}~\bibnamefont
		{Tanaka}},\ }\href {\doibase 10.1088/0034-4885/63/10/202} {\bibfield
	{journal} {\bibinfo  {journal} {Reports on Progress in Physics}\ }\textbf
	{\bibinfo {volume} {63}},\ \bibinfo {pages} {1641} (\bibinfo {year}
	{2000})}\BibitemShut {NoStop}%
\bibitem [{\citenamefont {Herrera}\ \emph {et~al.}(2010)\citenamefont
	{Herrera}, \citenamefont {Burset},\ and\ \citenamefont
	{Yeyati}}]{Herrera_2010}%
\BibitemOpen
\bibfield  {author} {\bibinfo {author} {\bibfnamefont {W.~J.}\ \bibnamefont
		{Herrera}}, \bibinfo {author} {\bibfnamefont {P.}~\bibnamefont {Burset}}, \
	and\ \bibinfo {author} {\bibfnamefont {A.~L.}\ \bibnamefont {Yeyati}},\
}\href@noop {} {\bibfield  {journal} {\bibinfo  {journal} {Journal of
		Physics: Condensed Matter}\ }\textbf {\bibinfo {volume} {22}},\ \bibinfo
{pages} {275304} (\bibinfo {year} {2010})}\BibitemShut {NoStop}%
\bibitem [{\citenamefont {Blonder}\ \emph {et~al.}(1982)\citenamefont
	{Blonder}, \citenamefont {Tinkham},\ and\ \citenamefont {Klapwijk}}]{BTK}%
\BibitemOpen
\bibfield  {author} {\bibinfo {author} {\bibfnamefont {G.~E.}\ \bibnamefont
		{Blonder}}, \bibinfo {author} {\bibfnamefont {M.}~\bibnamefont {Tinkham}}, \
	and\ \bibinfo {author} {\bibfnamefont {T.~M.}\ \bibnamefont {Klapwijk}},\
}\href {\doibase 10.1103/PhysRevB.25.4515} {\bibfield  {journal} {\bibinfo
	{journal} {Phys. Rev. B}\ }\textbf {\bibinfo {volume} {25}},\ \bibinfo
{pages} {4515} (\bibinfo {year} {1982})}\BibitemShut {NoStop}%
\bibitem [{\citenamefont {Tanaka}\ and\ \citenamefont
	{Kashiwaya}(1995)}]{Tanaka_1995}%
\BibitemOpen
\bibfield  {author} {\bibinfo {author} {\bibfnamefont {Y.}~\bibnamefont
		{Tanaka}}\ and\ \bibinfo {author} {\bibfnamefont {S.}~\bibnamefont
		{Kashiwaya}},\ }\href {\doibase 10.1103/PhysRevLett.74.3451} {\bibfield
	{journal} {\bibinfo  {journal} {Phys. Rev. Lett.}\ }\textbf {\bibinfo
		{volume} {74}},\ \bibinfo {pages} {3451} (\bibinfo {year}
	{1995})}\BibitemShut {NoStop}%
\bibitem [{Note1()}]{Note1}%
\BibitemOpen
\bibinfo {note} {The one-dimensional edge of a 2DTI, which is equivalent to
	the $\theta =0$ case considered here, presents a well-defined spin
	polarization. For the 2D surface state, the modes with $\theta \not =0$
	usually spoil this neat effect.}\BibitemShut {Stop}%
\bibitem [{\citenamefont {Reinthaler}\ \emph {et~al.}(2015)\citenamefont
	{Reinthaler}, \citenamefont {Tkachov},\ and\ \citenamefont
	{Hankiewicz}}]{Reinthaler_2015}%
\BibitemOpen
\bibfield  {author} {\bibinfo {author} {\bibfnamefont {R.~W.}\ \bibnamefont
		{Reinthaler}}, \bibinfo {author} {\bibfnamefont {G.}~\bibnamefont {Tkachov}},
	\ and\ \bibinfo {author} {\bibfnamefont {E.~M.}\ \bibnamefont {Hankiewicz}},\
}\href {\doibase 10.1103/PhysRevB.92.161303} {\bibfield  {journal} {\bibinfo
	{journal} {Phys. Rev. B}\ }\textbf {\bibinfo {volume} {92}},\ \bibinfo
{pages} {161303} (\bibinfo {year} {2015})}\BibitemShut {NoStop}%
\bibitem [{Note2()}]{Note2}%
\BibitemOpen
\bibinfo {note} {The orientation of the in-plane vector potential $\protect
	\mathbf { A }$ and, hence, that of the effective magnetization $\protect
	\mathbf { m }_{\parallel }$, is shifted from that of the magnetic field by
	$-\pi /2$.}\BibitemShut {Stop}%
\bibitem [{\citenamefont {Finck}\ \emph {et~al.}(2014)\citenamefont {Finck},
	\citenamefont {Kurter}, \citenamefont {Hor},\ and\ \citenamefont
	{Van~Harlingen}}]{Finck_2014}%
\BibitemOpen
\bibfield  {author} {\bibinfo {author} {\bibfnamefont {A.~D.~K.}\
		\bibnamefont {Finck}}, \bibinfo {author} {\bibfnamefont {C.}~\bibnamefont
		{Kurter}}, \bibinfo {author} {\bibfnamefont {Y.~S.}\ \bibnamefont {Hor}}, \
	and\ \bibinfo {author} {\bibfnamefont {D.~J.}\ \bibnamefont
		{Van~Harlingen}},\ }\href {\doibase 10.1103/PhysRevX.4.041022} {\bibfield
	{journal} {\bibinfo  {journal} {Phys. Rev. X}\ }\textbf {\bibinfo {volume}
		{4}},\ \bibinfo {pages} {041022} (\bibinfo {year} {2014})}\BibitemShut
{NoStop}%
\bibitem [{\citenamefont {Snelder}\ \emph
	{et~al.}(2015{\natexlab{b}})\citenamefont {Snelder}, \citenamefont {Stehno},
	\citenamefont {Golubov}, \citenamefont {Molenaar}, \citenamefont {Scholten},
	\citenamefont {Wu}, \citenamefont {Huang}, \citenamefont {van~der Wiel},
	\citenamefont {Golden},\ and\ \citenamefont {Brinkman}}]{Brinkman_2015}%
\BibitemOpen
\bibfield  {author} {\bibinfo {author} {\bibfnamefont {M.}~\bibnamefont
		{Snelder}}, \bibinfo {author} {\bibfnamefont {M.}~\bibnamefont {Stehno}},
	\bibinfo {author} {\bibfnamefont {A.}~\bibnamefont {Golubov}}, \bibinfo
	{author} {\bibfnamefont {C.}~\bibnamefont {Molenaar}}, \bibinfo {author}
	{\bibfnamefont {T.}~\bibnamefont {Scholten}}, \bibinfo {author}
	{\bibfnamefont {D.}~\bibnamefont {Wu}}, \bibinfo {author} {\bibfnamefont
		{Y.}~\bibnamefont {Huang}}, \bibinfo {author} {\bibfnamefont
		{W.}~\bibnamefont {van~der Wiel}}, \bibinfo {author} {\bibfnamefont
		{M.}~\bibnamefont {Golden}}, \ and\ \bibinfo {author} {\bibfnamefont
		{A.}~\bibnamefont {Brinkman}},\ }\href@noop {} {\  (\bibinfo {year}
	{2015}{\natexlab{b}})},\ \bibinfo {note} {arXiv:1506.05923}\BibitemShut
{NoStop}%
\end{thebibliography}
\end{document}